\documentclass[letterpaper,twocolumn,10pt]{article}
\usepackage{usenix}

\usepackage{courier} 

\usepackage{tikz}
\usepackage{amsmath}

\usepackage{filecontents}

\usepackage{listings}
\usepackage{xcolor}
\usepackage{graphicx}
\usepackage{subfig}
\usepackage[textsize=tiny,disable]{todonotes}
\usepackage{amssymb}

\usepackage[normalem]{ulem}

\colorlet{punct}{red!60!black}
\definecolor{background}{HTML}{EEEEEE}
\definecolor{delim}{RGB}{20,105,176}
\colorlet{numb}{magenta!60!black}

\lstdefinelanguage{json}{
    basicstyle=\footnotesize\normalfont\ttfamily,
    numbers=left,
    numberstyle=\scriptsize,
    stepnumber=1,
    numbersep=8pt,
    showstringspaces=false,
    breaklines=true,
    frame=lines,
    backgroundcolor=\color{background},
    literate=
     *{:}{{{\color{punct}{:}}}}{1}
      {,}{{{\color{punct}{,}}}}{1}
      {\{}{{{\color{delim}{\{}}}}{1}
      {\}}{{{\color{delim}{\}}}}}{1}
      {[}{{{\color{delim}{[}}}}{1}
      {]}{{{\color{delim}{]}}}}{1},
}

\begin{document}
\setlength{\marginparwidth}{1.55cm}
%-------------------------------------------------------------------------------

%don't want date printed
\date{}

\makeatletter
\newcommand{\printfnsymbol}[1]{%
  \textsuperscript{\@fnsymbol{#1}}%
}

\title{\Large \bf Hey Google, What Exactly Do Your Security Patches Tell Us?\\A Large-Scale Empirical Study on Android Patched Vulnerabilities}

%for single author (just remove % characters)
\author{
{\rm Sadegh Farhang \thanks{Sadegh Farhang and Mehmet Bahadir Kirdan equally contributed to this work.}}\\
{\rm Pennsylvania State University}\\
{\rm \texttt{smf5604@psu.edu}}
\and
{\rm Mehmet Bahadir Kirdan \printfnsymbol{1}}\\
{\rm Technical University of Munich}\\
{\rm\texttt{bahadir.kirdan@tum.de}}
\and
{\rm Aron Laszka}\\
{\rm University of Houston}\\
{\rm \texttt{alaszka@uh.edu }}
\and
{\rm Jens Grossklags}\\
{\rm Technical University of Munich}\\
{\rm \texttt{jens.grossklags@in.tum.de}}
% copy the following lines to add more authors
% \and
% {\rm Name}\\
%Name Institution
} % end author

\maketitle

%-------------------------------------------------------------------------------
\begin{abstract}
Android has the largest market share among smartphone platforms worldwide with more than one billion active devices. 
%\Aron{text mentions no software vendor, so ``other software vendors'' sounds a bit strange to me}
Like other platforms, security patches play a pivotal role in keeping Android devices safe from the exploitation of known vulnerabilities. Previous research efforts have documented many attacks, vulnerabilities, and defenses in the Android ecosystem. However, no previous work has  
studied Android vulnerabilities and their implications on consumers, public vulnerability disclosure,
and the Android ecosystem together. 

In this paper, we perform a comprehensive study of 2,470 patched Android vulnerabilities that we collect %\change{scrape}{collect} 
from different data sources such as Android security bulletins, CVEDetails, Qualcomm Code Aurora, AOSP Git repository, and Linux Patchwork. In our data analysis, we focus on determining the affected layers, OS versions, severity levels, and common weakness enumerations (CWE) associated with the patched vulnerabilities. Further, we assess the timeline of each vulnerability, including discovery and patch dates.

We find that (i) even though the number of patched vulnerabilities changes considerably from month to month, the relative number of patched vulnerabilities for each severity level remains stable over time, 
(ii) there is a significant delay in patching vulnerabilities that originate from the Linux community or concern Qualcomm components, even though Linux and Qualcomm provide and release their own patches earlier, 
(iii) different AOSP versions receive security updates for different periods of time, 
(iv) for 94\% of patched Android vulnerabilities, the date of disclosure in public datasets is not before the patch release date, (v) there exist some inconsistencies among public vulnerability data sources, e.g., some CVE IDs are listed in Android Security bulletins with detailed information, but in CVEDetails they are listed as unknown, (vi) many patched vulnerabilities for newer Android versions likely also affect older versions that do not receive security patches due to end-of-life.
\end{abstract}
%-------------------------------------------------------------------------------

%-------------------------------------------------------------------------------
\section{Introduction}
\label{sec:Intro}

Modern mobile phones have become an integral part of our lives. They are the central information hub for many users, harboring deeply personal information, and are also the conduit for many economically relevant transactions, such as mobile banking or healthcare. As a result, the focus of cybercriminals and other attackers has also shifted towards this context.
%Due to their ubiquity and applications, %\change{the ubiquity of them and their applications}{their ubiquity and applications}, 
%we use smartphones for various purposes ranging from financial, personal, health-related, 
%\Aron{are you using Oxford comma?} 
%and so forth\Mehmet{etc.?}. They are a \Aron{\url{https://www.thefreedictionary.com/reservoir} let's replace ``reservoir''}reservoir of invaluable information, which lures many attackers to compromise users' smartphones. 
Thus, it is essential to keep smartphones secure with particular emphasis given to central key functions such as %\change{smartphones}{them} 
the operating system (OS), like Android and iOS. 
%Security of smartphones consists of different components like operating system (OS) and application security

In our work, we %\change{only focus}{focus only} 
focus on Android, the mobile operating system developed by Google and released under open-source licenses as the Android Open Source Project (AOSP). The first commercial Android device was launched in September 2008, and  Android now has the largest market share among smartphone platforms worldwide with more than one billion active devices~\cite{market_share}. 

%Nowadays, smartphones are an integral part of our lives. Due to ubiquity of smartphones and their wide applications, we use smarphones for wide range of purposes ranging from financial, health, personal, and so forth. Smartphones are reservoir of invaluable information which lures many attackers to compromise users' smartphones. Thus, it is essential to keep smartphones secure. In order to keep smartphones secure, one of the most important parts is the operating system (OS), like Android and iOS. 
%Security of smartphones consists of different components like operating system (OS) and application security
%Here, we focus only on Android, the mobile operating system developed by Google and released under open-source licenses as the Android Open Source Project (AOSP). The first commercial Android device was lunched in September 2008 and now it has the largest market share among smartphone platforms worldwide with more than one billion active devices~\cite{market_share}. 

Similar to other software vendors, Google provides (monthly) security bulletins, which contain the details of patched vulnerabilities affecting multitudes of Android devices. %\Aron{grammar is ambiguous, clause ``starting from August...'' could apply to ``contain'' instead of ``provides''} 
Google has been publishing its Android security bulletins starting from August 2015 to the present~\cite{Android_Security_Bulletins}. Similar to Google Android security bulletins, some other vendors initiated their own security bulletins. For instance, Samsung and LG started to publish security bulletins in October 2015 and in May 2016, respectively~\cite{Samsung_Security_Bulletin, LG_Security_Bulletin}. %\Aron{is Qualcomm listed in a separate sentence because it is not a phone vendor? if so, then it might be worth to mention this}  Qualcomm also publishes security bulletins starting from December 2017~\cite{Qualcomm_Security_Bulletin}.

%But, why Google have not tried yet to catch up and provide the data before that time. 

%Like other software vendors, Google provides monthly patch updates to vulnerabilities that have been discovered to keep Android secure.  
%Further, Google started publishing Android security bulletins in August 2015 and continue to do that so far~\cite{Android_Security_Bulletins}. A security bulletin publishes monthly and it contains the details security vulnerabilities affecting Android devices being patched in that month. 
%But, why Google have not tried yet to catch up and provide the data before that time. 
%Following Google Android security bulletins, some vendors started their own security bulletins. For instance, Samsung started in October 2015 and LG started in May 2016.

There has been extensive work to find vulnerabilities in the Android ecosystem, such as~\cite{shabtai2010securing, yang2018study,ren2015towards}, in addition to enhancing %\change{make}{making} 
its security, e.g., ~\cite{ren2017windowguard, enck2009understanding}. %\Aron{don't get what?}\textcolor{red}{DON'T GET IT}. 
Despite these efforts, %\change{ to make a more secure environment}{}, 
we observe %\Aron{\url{https://www.thefreedictionary.com/realize}}\change{realize}{observe} 
an increase in the number of attacks and vulnerabilities. 
%\Aron{remove ``According to [X],'' (or replace with, e.g., ``According to a recent study,'') and add citation [X] to the end of the sentence}According to~\cite{TotalNoVulner}, 
In 2016, the total number of publicly disclosed %\Aron{publicly disclosed vulnerabilities?} 
vulnerabilities for all platforms (i.e., all vendors and product versions) reached 6,447. However, %\change{it raised}{the number rose}
the number increased to 14,714 in 2017 and maintained this upward trend in 2018 by reaching %\change{to}{} 
16,555~\cite{TotalNoVulner}. %This case  also true for Android related vulnerabilities. 
%\Aron{good question, I was wondering about the same}.
%Therefore, it is crucial to understand why the trend is increasing despite significant efforts to improve security.
%essential to focus not only %\change{not only focus}{to focus not only} 
%on vulnerability detection and defenses, \Aron{I would say that it is crucial to understand why the trend is increasing despite significant efforts to improve security} but also %\change{}{to} 
%to investigate and understand the general trend of Android vulnerabilities over time.  

%There has been extensive work to find vulnerabilities in Android ecosystem, such as~\cite{shabtai2010securing, yang2018study,ren2015towards}, in addition to make it more secure, e.g., ~\cite{ren2017windowguard, enck2009understanding}. Despite these efforts to make more secure environment, we are witnessing an increase in number of attacks and vulnerabilities. According to~\cite{TotalNoVulner}, in 2016, the total number of vulnerabilities are 6447. The total number of vulnerabilities have increased tremendously in 2017 and 2018. The total number of vulnerabilities are 14,714 an 16,555 in 2017 and 2018, respectively. This case is also true for Android related vulnerabilities. Therefore, it is essential to not only focus on vulnerability detection and defenses, but also investigate and understand the general trend of Android vulnerabilities over time. 

In this paper, %\Aron{move this clause to the end of the sentence} 
we perform the most comprehensive analysis of Android vulnerabilities, to the best of our knowledge. We create a dataset containing 2,470 patched vulnerabilities with their detailed information by scraping %\change{}{multiple} 
multiple data sources, e.g., Android security bulletins, CVEDetails, Google Gits. This dataset enables us to perform %\Aron{different compared to what?} 
multiple analyses to better understand security in the Android ecosystem. %\Aron{this sentence is a bit vague, it could be removed since it is followed by a more detailed list of findings} In particular, we focus on the timing of vulnerabilities, affected components, OS versions. 
In the following, we summarize our findings. 
%\Aron{I suggest making this an itemized list to give a nice overview to the reader of the main contributions} 
\begin{itemize}
    \item We document the expected behavior that Google does not provide updates for all affected OS versions due to end-of-life (EOL), which is the point in time at which a company ceases to create any further updates (including patches) for a given OS version. %\Aron{define acronym (I know it's trivial, but some reviewers might complain} 
    However, many patched vulnerabilities are likely common among different OS versions. Therefore, OS versions that reached their EOL are vulnerable to many vulnerabilities, which are patched for newer versions. This practice puts consumers of outdated operating systems at risk. %We investigate this issue in detail in Section~\ref{subsub:InConsistent} and Section~\ref{subsub:common}.
    %\Aron{when I read this first, it was really unclear what ``that'' refers to (might be better to change to ``this'' or restructure the sentences)} that, this practice puts many outdated OS versions at risk.
    
    \item We find that different OS versions receive security updates for different periods of time. For example, the difference in introduction dates of versions 4.4 and 4.4.4 is about 8 months; however, version 4.4 stopped receiving updates 22 months before version 4.4.4 reached EOL. This difference in duration of receiving security support differentiates among its consumers. Consumers who update their OS are rewarded by receiving security support for a longer period of time. 
    %Aron{there are a lot of short sentences in a row, it might improve readibility to merge some} 
    %Some versions receive security updates for a very short period of time.
    
    \item Some Android vulnerabilities originate from other resources, e.g., Qualcomm and Linux. We observe a significant delay from the Android security team to provide patches for these vulnerabilities, while Qualcomm and Linux provide and release their own patches earlier. The average delays for Qualcomm and Linux are 307.8 and 324.9 days, respectively.
    %Some Android vulnerabilities %\change{are originated}{originate} 
    %originated from other resources, e.g., %\Aron{did you mean ``e.g.''?}i.e., 
    %Qualcomm and Linux. We observe a significant delay from the Android security team to provide patches for those vulnerabilities. The average delay for Qualcomm and Linux are 307.80 and 324.86 days, respectively. 
    %\Aron{it would be good to compare these numbers to the average patching time for other vulnerabilities}.
    
    \item We study the patch release dates, fix-commit dates, and public repository disclosure dates for patched vulnerabilities. We find that for the majority, i.e., 94\%, the patch release date occurs before or in the same month as the public disclosure date, which can be considered a prudent, secure practice. We discuss whether disclosure that predates fixing and patching places consumers at risk. %Even the rest is negligible, it is an insecure practice that puts consumers at risk. %(Section~\ref{sub:delay}).
    
    \item We determine 
    %\Aron{move ``by using SZZ algorithm'' to the end of the sentence, current structure can be confusing} by using \Aron{what is this?}SZZ algorithm that represents 
    the time difference between the introduction of the first line of code associated with a vulnerability and the publication time of a security bulletin, which is called \textit{maximum vulnerability lifetime}.
    %by using SZZ algorithm~\cite{SZZ}. The average time is equal to %\Aron{number of decimal digits is inconsistent (see nubmers 307.80 and 324.86 above)} 
    We find that the average maximum vulnerability lifetime is about 1350 days, which provides an understanding of how good the state of the art security tools are in terms of detecting and fixing Android vulnerabilities.  %and 882 days for maximum and minimum vulnerability lifetimes, respectively. 
    
    \item Even though the monthly number of patched vulnerabilities for each severity level\footnote{Google uses the name \textit{severity ratings}. Here, we use severity levels instead of severity ratings.} changes considerably, we find that the relative number of patched vulnerabilities for each severity level has a similar distribution mean each year.  
    %\Aron{it would be good to simplify this sentence} which are grouped based on their severity levels
    %\Aron{is this just a matter of naming? if some, make this clear}
    %\footnote{Google uses severity ratings. Here, we use severity levels instead of severity ratings.} have the same distributions.
    
    \item We observe some inconsistencies between Android security bulletins and CVEDetails. Some CVE IDs are listed in Android Security bulletins with detailed information, while in CVEDetails those CVEs often remain unknown. 
    %CVE IDs in Android security bulletins with detailed information the existence of some \textit{unknown CVE IDs} in CVEDetails website, while Android security bulletins provide \change{detail}{detailed} information about them\Aron{I'm a little confused by this sentence, are the CVE IDs listed on Android sec bulletin?}. We discuss them in detail in Section \Aron{is the section missing or just the reference?}\textcolor{red}{add}.

\end{itemize}

The remainder of this paper is organized as follows. In Section~\ref{sec:RelWork}, we discuss related work. In Section~\ref{sec:RQ}, we %\change{propose}{pose} 
pose our research questions. We describe our data collection methodology and our dataset in Section~\ref{sec:data_set}, which is followed by the presentation of our results in Section~\ref{sec:results}. We discuss our findings in Section~\ref{sec:Discuss} and limitations in 
Section~\ref{sec:Limit}. We offer concluding remarks in Section~\ref{sec:conclusion}.

%This is what Google has provided in its Security Bulletin as the main creator of Android. Therefore, it is probably the most reliable resource to investigate the overall trend of Android security ecosystem even though there are many missing and unclear issues. As a result, we avoid adding any other information or inferred some other things from other ways. We try to investigate what these security bulletin conveys by looking just to itself about Android security ecosystem. 

%The number of vulnerabilities are increasing these days. Based on https://www.cvedetails.com/browse-by-date.php, in 2016, the total number of vulnerabilities are 6447. The total number of vulnerabilities have increased tremendously in 2017 and 2018. The total number of vulnerabilities are 14,714 an 16,555 in 2017 and 2018, respectively. 

%The paper that introduced the notion of CVE~\cite{mann1999towards}

%\cite{linares2017empirical}

%\cite{android_build_numbers}

%\cite{android_versions}

%\cite{android_stack}
%-------------------------------------------------------------------------------

%-------------------------------------------------------------------------------
\section{Related Work}
\label{sec:RelWork}

%\Aron{this has been stated multiple times already}Android has the largest market share among mobile platforms over the last years~\cite{Android_Market_Share}. 
%\Aron{what does ``these mobile devices'' refer to? there are no mobile devices mentioned in the previous sentence} 
Mobile devices contain lots of sensitive and private information. As a result, security should be an integral part of the Android ecosystem.
Significant research efforts have been spent on investigating attack techniques on Android, finding vulnerabilities, and designing a more secure infrastructure in Android~\cite{vidas2011all, xing2014upgrading, enck2009understanding, shabtai2010securing}. %\change{Like}{Similar to} 
Similar to other platforms, within the Android ecosystem  %\change{}{its} 
 security patches are developed and issued in a regular fashion to maintain device security. Contrary to previous works, we follow a different approach to investigate the overall practices related to vulnerability disclosure and security patching in the Android ecosystem by gathering and analyzing Android security bulletins. While the previous research on Android-related vulnerabilities and their implications is sparse~\cite{linares2017empirical}, software updates, as well as Android security, have been investigated from different perspectives~\cite{jo2017effect, gkantsidis2006planet, duebendorfer2009silent}.

\subsection{Android-Related Vulnerabilities}
Most closely related to our work, Linares-V{\'a}squez et al.~\cite{linares2017empirical} conducted an empirical study based on the collection of 660 Android-related vulnerabilities mined from Android security bulletins, CVEDetails and XML feeds provided by NVD. They investigated three issues. First, they studied the CWE hierarchies and the types of vulnerabilities affecting Android. Second, they studied the Android layers affected by vulnerabilities. Third, they investigated the time intervals between the introduction date of the vulnerability and its fix date. Note that they mined the patched vulnerabilities up to November 2016. On the contrary, we collect 2,470 patched vulnerabilities in Android from August 2015 up to January 2019, which is more comprehensive. 
Moreover, our study not only covers all of their analysis with a more comprehensive dataset, but also studies new issues. %from analyzing the maximum and minimum survivability lifetimes, severity levels and CWEs of the vulnerabilities,
These new investigations include, but are not limited to, the duration of security support for different versions and the delay in patching vulnerabilities originating from Linux and Qualcomm.  
%release dates, update duration of the versions as well as the comparison of updated and affected versions of AOSP. Moreover, we analyze not only the time sequence of the public disclosure date, fix-commit date and patching date of vulnerabilities but also the time period of retrieving vulnerability patches from other vendors such as Qualcomm and Linux.
%\Mehmet{Still not sure again to say what we did here?}
%\Aron{will other references be included? if not, I suggest merging this subsection into another one}

\subsection{Software Updates}

Several works investigate the delivery and installation of patches across devices. Nappa et al.~\cite{nappa2015attack} analyzed the life cycle of %\change{client applications' vulnerabilities}{vulnerabilities in client applications} 
vulnerabilities in client applications by observing the deployment of patches on users' devices. They used the Worldwide Intelligence Network Environment (WINE) as their data source and found that the patching rates differ among applications. 
%\Aron{how effective the update mechanisms are in practice?} how the update mechanisms are effective in practice.
Mathur and Chetty~\cite{mathur2017impact} studied the issue of semi-automatic updates in Android mobile devices and the impact of user experiences on update behavior. Taking a similar research focus, Edwards et al. reported that purely automatic updates are subject to failure~\cite{edwards2008security}. Considering the issues related to full security automation, %\Aron{human behavior?}
human behavior and vendor policies (in our case, we mostly focus on Google) remain an integral part of the security decision-making process~\cite{akhawe2013alice, sunshine2009crying}. 

From the user perspective, there are several works that %\change{}{that} 
study the role of humans regarding software updates and upgrades. Vaniea and Rashidi~\cite{vaniea2016tales} found six stages that users go through during their software updates. These six stages are awareness, deciding to update, preparation, installation, troubleshooting, post-state of the update. 
Farhang et al.~\cite{farhang2018take} conducted a survey to better understand the relevant factors for upgrade decisions in Microsoft operating systems by recruiting 239 participants. They studied how users perceive privacy issues associated with OS upgrade decisions, and whether security constitutes a significant decision-making factor. %differentiated between the terms update and upgrade \Aron{why and how?}. They also studied the underlying reasons for upgrade practices in Microsoft Windows operating systems. 
%also did a user survey among 239 Microsoft Windows users to analyze their upgrade practices.

Frei et al.~\cite{frei20080} also investigated the life cycle of Microsoft and Apple vulnerabilities by defining a new metric called %\Aron{this is the only thing in bold font so far, which draws a lot of attention to it, I suggest making it only italic} 
\textit{0-day patch rate}. They defined it as the number of vulnerability patches that a vendor releases %\change{at}{} 
when the vulnerabilities are publicly disclosed. Using this metric, they compared Apple with Microsoft in terms of their security performance and concluded that while Apple shows an ascending trend, Microsoft is more stable than Apple with regard to the average number of unpatched vulnerabilities.
%successful to keep the average number of unpatched vulnerabilities below 20 while Apple is unstable in this case. 
Further, Li and Paxson~\cite{li2017large} scraped 3,000 vulnerabilities that belong to 682 different open-source software projects and studied the patch development life cycle.

Shahzad et al.~\cite{shahzad2012large} conducted %\change{did}{conducted} 
an exploratory measurement study of %\Aron{46,310?} 
46,310 vulnerabilities disclosed from 1988 to 2011 concerning different software vendors. %\change{The}{Their} main goal \change{is}{was} 
Their main goals were to analyze disclosure trends, as well as the evolution of CVSS scores and so-called vector metrics such as confidentiality impact, access complexity, and availability impact. One of their findings was that the annual number of disclosed vulnerabilities has stopped increasing since 2008, which is opposite to what we are observing now in the Android context. %\Aron{what did they find?}

\subsection{Economics of Software Updates and Android Security}

Arora et al.~\cite{arora2010empirical} investigated the relation between software vulnerability disclosure and patch release time. They found that disclosure accelerates patch release, and open source vendors are quicker in releasing the patches. Alhazmi and Malaiya~\cite{alhazmi2005modeling} focused on different vulnerability discovery models in operating systems and compared the results of the models with each other by performing statistical tests. One of the limitations of their model was that it did not differentiate among vulnerabilities with different severity levels. %\Aron{any interesting findings?}

Jo~\cite{jo2017effect} proposed a model for examining competition intensity in the context of patching security vulnerabilities for free-of-charge software products. She found that an increase in market concentration improves vendor response in patching vulnerabilities. To test the model, she focused on the web browser market. Farhang et al.~\cite{farhang2017economic} studied the issue of competition in a different domain, i.e., the competition in the Android ecosystem and customization of different vendors' offerings which may result in security issues. To solve that issue, they proposed a regulatory fine model, which incentivizes a certain level of investment in security while decreasing the price of products. 

A set of studies also investigates the ecosystem around so-called bug bounty platforms, which host vulnerability discovery programs for different companies and offer (monetary) incentives for white hat hackers to participate \cite{Laszka16,maillart2017given,zhao2015empirical}. While a number of companies offering IoT and mobile services participate in bug bounty programs, Android is not a primary focus of these platforms.

%\cite{ruohonen2018bug}

%Edwards et al.~\cite{edwards2016hype} analyzed the trend of data breaches by using the Bayesian generalized linear models. 

%Since the economical side of Android ecosystem is also crucial, Farhang et al. \cite{farhang2017economic} developed a model to utilize the concepts of game theory and product differentiation to capture the competition involving two vendors customizing the AOSP platform.

%-------------------------------------------------------------------------------

%-------------------------------------------------------------------------------
\section{Study Design}
\label{sec:RQ}
%\textcolor{red}{It might be better to move that part to Intro.}

To have a better and systematic understanding of Android security vulnerabilities and patches from different perspectives, we focus on the following five research questions.
%Based on these questions, the process of data generation and analyses will be explained in the following sections.

\textbf{RQ1}: \textbf{\textit{How have severity distribution and root causes of 
%\Aron{maybe mention what aspects of vulnerabilities you study, e.g., severity distribution, patching/discovery rate, root causes} 
patched Android vulnerabilities evolved over time?}}
%What is the overall picture on Android security patches?}}

Google has published 42 monthly Android security bulletins since August 2015 until the time of data collection, 
%\change{writing of this paper}{data collection}, 
January 2019. Each of these monthly security bulletins consists of several Android vulnerabilities and their patch details.
%that Google has provided patches for them till that month. 
%\Aron{for what reason?}
%For this reason, in order 
To have a better understanding of these vulnerabilities, it is essential to investigate %\change{the overall trend of them}{their overall trend} 
their overall distribution over time.
%general picture of them. 
To achieve this, %\change{that}{this}, \change{first, we}{we first} 
we first investigate the severity levels of patched Android vulnerabilities.%, which 
%\change{ that are patched so far}{}\change{. This}{, which} 
%gives us a better understanding of how severe Android vulnerabilities are. 
We also study the trend in the number of patched vulnerabilities. 
%\Aron{this is not clear to me, are you saying simply that you study the trend in the number of patched vulnerabilities?} improvements in Android ecosystem in terms of the number of patched vulnerabilities. 
Next, we classify the vulnerabilities based on their Common Weakness Enumerations (CWE). This enables us to recognize what the %\change{main reasons for}{primary causes of}  
common causes of vulnerabilities are.
%and to identify which of these \Aron{by weakness, do you mean reason/cause? if yes, then wouldn't the primary ones be the common ones anyway?} weaknesses are common in Android vulnerabilities. 
%\Aron{this sentence is important! you need to show that the results presented in this paper will have benefit / impact (e.g., guiding developers to focus on the important issues); these points should be mentioned in the abstract / intro, otherwise the reader might think that the results are interesting but pointless} 
These results can help Android developers to identify problematic practices and areas that necessitate heightened attention to prevent such common causes of vulnerabilities. 
%This also gives Android developers key issues that they need to focus when they are building 
%As a result, we can help those who want to contribute to the Android ecosystem for mitigating these weaknesses.

%Second, we classify vulnerabilities based on their Common Weakness Enumerations (CWE). This enables us to know what the sources of vulnerabilities are and which are more common in Android which can also help developers by giving them some warning about which type of vulnerabilities they are more prone to in Android ecosystem

%Firstly, all patched vulnerabilities so far were indicated by grouping them their severity levels monthly. By having this analysis, the overall performance of Android security patches can be seen. Secondly, since each of these vulnerabilities has their own unique types, they were grouped by looking their Common weakness Enumeration (CWE) types. Having a solid notion of grouping them by their types could help other researches as well as the developers for understanding them what kind of weaknesses are playing major roles.

\textbf{RQ2}: \textbf{\textit{Is the duration of security %\Aron{security support?} updates 
support equal for different AOSP versions?}}
%What is the effect of ending support for old versions on consumers' security?}}
%\textcolor{red}{DOES THIS QUESTION FIT OUR ANALYSIS?}
%How many of AOSP versions are affected and how many of them are updated?}}

Google has 28 Android API levels and has released 63 different AOSP versions since 2008~\cite{All_AOSP_Versions}. Most of these versions are still used by consumers. For example, at the time of writing 
%\change{this paper}{} 
(January 2019), the market shares of versions %\change{}{versions} 
Froyo and Ginger are 0.02\% and 0.25\%, respectively~\cite{mobile_android_version_share_worldwide}. However, Google stops providing security updates for each version after some time. For instance, Google's policy for Pixel devices is as follows: ``Pixel phones get security updates for at least 3 years from when the device first became available on the Google Store, or at least 18 months from when the Google Store last sold the device, whichever is longer. After that, we cannot guarantee more updates''~\cite{update_nexus_device}.
%Therefore, it is obvious that Google does not provide security patch updates for older versions even if they are affected by any vulnerability. 
%\Aron{why do you expect this?} The expectation is that 
%Different AOSP versions should receive the security patch updates for the same time duration.
In this question, we investigate %\change{the release date of each version and find for}{} 
how long each Android %\change{that}{each Android} 
version receives security patch updates. As a result of this investigation,  %\change{question}{investigation}, 
we observe that Google has provided security patch updates for some versions for shorter periods of time than for other versions. For example, the difference in introduction dates of version 4.4 and version 4.4.4 is about 8 months. However, version 4.4 stopped receiving updates 22 months earlier. %\Aron{any numbers that could be easily included here?}. 
We also investigate the patched AOSP versions and the affected AOSP versions over time. We can demonstrate that the Android security team does not provide security patch updates for all affected versions because older versions have reached their %\Aron{acronym EOL should be defined when first used}
end-of-life. Considering that most AOSP versions are still used by consumers, this practice leaves %\change{are left}{leaves} 
many devices and consumers unprotected. 

%To answer this question, Android security bulletins only provided updated AOSP versions for some vulnerabilities. To have affected versions,  

%Each of these AOSP versions was used by several end users. Some of the older versions have been still using even though there are updated versions released by Google. Although there are updated AOSP versions that were specified on Android security bulletins, there are also affected AOSP versions. It may be an issue that some of the affected versions may not be updated. Although there could several factors such as the security patch policies, this issue are still crucial to understand the performance of the updating the affected versions. Such kind of analysis could give opinions about how many of AOSP versions could be still vulnerable. Given this motivation, the updated AOSP versions that have been publishing for each vulnerability patch are used as well as the affected versions that were published on CveDetails \cite{Cve_Details}.

\textbf{RQ3}: \textbf{\textit{How long does it take for Google to 
%\Aron{what does handle mean? include downstream?} 
patch vulnerabilities originating from other resources?}}

%How long does an external vulnerability patch last for publishing on an Android security bulletin?

%How long does it take for AOSP to handle vulnerabilities of other resources?

%What is the role of entities other than Google in patch timeline?}}
%How long does importing external vulnerability patches into AOSP take?}} 

Besides vulnerabilities that are specific to Android, there are vulnerabilities that affect other related key software building blocks. %\change{others which affect the parts that do not only belong to Android}{vulnerabilities that affect other software as well}.
For instance, kernel development is driven by the Linux community, which patches kernel-related vulnerabilities separately. Qualcomm components are used in Android and
%\Aron{I think that Qualcomm is a company, not an ``entity in Android''} entity in Android and 
Qualcomm also patches their Qualcomm-related vulnerabilities by itself. In our Android dataset, we find 1092 Qualcomm-related and 213 kernel-related patched vulnerabilities, and they all have references that point %\change{navigate}{point} 
to their corresponding repositories. These repositories provide detailed %\change{detail}{detailed} 
information about the patch including the fix-commit date. 
We investigate the time gaps between the last fix-commit date and the published date on Android security bulletins. Thus, we can calculate how long it takes for Google to publish Qualcomm and Linux patched vulnerabilities in Android security bulletins.
%\change{Qualcomm and Linux patched vulnerabilities take for Google to publish them on}{it takes for Google to publish Qualcomm and Linux patched vulnerabilities in} Android security bulletins. 

%Since there are 1092 Qualcomm-related, 213 kernel-related vulnerabilities and they have published references for the patches, we investigate the patching time of Qualcomm and Kernel and publishing time in Android security bulletins. By analysing them, we can see the time differences between Google and Qualcomm and Linux. SHOULD BE READ ONE MORE TIME

\textbf{RQ4}: \textbf{\textit{Are vulnerabilities in Android patched before their public disclosure?}}
%\change{Does a vulnerability in Android patch before its}{Are vulnerabilities in Android patched before their} public disclosure?}}
%Do all vulnerabilities follow the same time sequence for patching?}}
%How long does a vulnerability patch last from its publicly disclosure to bulletin time?

Google publishes the patched vulnerabilities on its monthly security bulletins with an exact bulletin publication date. 
%dates for each of  \Aron{``them'' is ambiguous} them. 
%\Aron{lot of short sentences in a row} 
This date can also be considered as patch release date. For example, in the January 2019 security bulletin, it is remarked that
%\Aron{please simplify these opening clauses} According to Google in one of its security bulletins, e.g., in January 2019~\cite{Android_Security_Bulletin_2019_01}, 
``the Android security bulletin contains details of security vulnerabilities affecting Android devices. Security patch levels of 2019-01-05 or later address all of these issues''~\cite{Android_Security_Bulletin_2019_01}.
As a result, we use a term called \textbf{\textit{patch release date}} that indicates the publishing date of a patched vulnerability on its security bulletin as well as the patch release for that vulnerability. 
We also use \textbf{\textit{public disclosure date}} to specify when vulnerability details are publicly available in public datasets like CVEDetails. 
%In other words, the time when everyone can read its details at. Therefore, we call it as \textbf{\textit{disclosure date}}. 
Considering that some of the patched vulnerabilities have references that navigate to Git %\change{git}{Git} 
repositories, we also retrieve the last commit dates that fix the vulnerability and we refer to it as \textbf{\textit{last commit date}} or short \textbf{\textit{commit date}}. 
In an ideal and secure situation, the ordering of these three notions should typically be as follows. The last commit date should be first and the public disclosure time should not be earlier than the patch release date.  
%Then a vulnerability patch should be published on a security bulletin. Then the vulnerability details should be publicly disclosed so that everyone can reach the details of it.
We investigate the actual ordering of these three notions in Android and their respective frequencies. For example, we can evaluate whether a patch is released no later than the public disclosure date of the corresponding vulnerability. 
%Given this hypothesis, we investigate the time sequences of these three times to see whether they are in the order as we claim.

%Google publishes the vulnerability patch on its security bulletins. Hence, we have bulletin dates for each vulnerability dates. we propose a term called bulletin date that indicates the publishing date of a vulnerability patch on its security bulletins. Publicly disclosure means, e.g.disclosure date when a vulnerability detail is publicly available. Last fixing-commit date is the date when a last commit that patches a vulnerabilitty is made. In an ideal timeline, first the last commit should be made. The bulletin date should come after the last fixing-commit date. Disclosure date should be the last date. Given this hypothesis, we investigate the day differences of these three times to see whether they are in the order like we claim.

\textbf{RQ5}: \textbf{\textit{How long does it take to patch a vulnerability?}}
%\change{}{it take to patch} a vulnerability\change{ take to patch}{}?}}
%How long does it take to make fixes for Android vulnerabilities?
%What are the survivability days of patched vulnerabilities?}}

We use a term called 
%\Aron{hasn't such a concept been used in prior work? I find it very surprising that no one used this metric before (if they did, then don't claim that you propose it)}
\textbf{\textit{vulnerability lifetime}} that indicates the time difference between a vulnerability introduction and its publishing time on the bulletin. The aim of this research question is to find the distribution of vulnerability lifetime for vulnerabilities with different severity levels. This gives us a better implicit understanding of how good the state of the art security tools are in terms of detecting and fixing Android vulnerabilities. %Furthermore, the vulnerability lifetime analysis shows that 
Likewise, this analysis provides an indication whether we invest enough resources to reduce the vulnerability lifetime considering the huge market share and impact of Android devices in our daily lives. 

%In general, a vulnerability is patched by changing several lines of code on some files. Note that these changes in the code are corresponding to the lines that have created the vulnerability. 

%Each of vulnerability patching is done by changing several lines on several files. At this point, it is important that these changed lines are the ones that causes the vulnerability. However, these lines that are causing the vulnerabilities were implemented before. This means that a vulnerable can be exist even before it is not publicly disclosed. Therefore, the time difference between the time when may-cause-vulnerability lines were implemented in and the time when a vulnerability patch was indicated on a security bulletin in. To do this analysis, the bug-fixing links that have been publishing for each vulnerability patches were used. 
%Hence, the questions of how many days a vulnerability can survive 
%and what kind of possible techniques or tools can be used in order to indicate a vulnerability can be discussed. 

%------------------------------------------------------------------------------

%-------------------------------------------------------------------------------
\section{Data Collection Methodology}
\label{sec:data_set}

In this section, we describe the collection process for the dataset to provide answers to the research questions posed in Section~\ref{sec:RQ}. The research questions shape our data collection methodology and indicate what information we need to provide answers.
%\change{answer those questions}{provide answers}. 

Our main source of data are the Android security bulletins for the date range
%\change{}{of data}
 from August 2015 to January
%\change{of them}{bulletins} 
%\Aron{download?}
2019~\cite{Android_Security_Bulletins}. Note that August 2015 is also the start date for publication of the security bulletins. However, the security bulletins do not provide all our required information regarding Android vulnerabilities, like CWE types and public disclosure dates. Therefore, we also crawl %\Aron{crawl? use?} 
%\change{We \change{do}{performed} our first scraping of both CVEDetails and the bulletins in mid-October 2018. However, due to several updates on both Android security bulletins and CVEDetails website, we conduct our second scraping at the end of January 2019. Hence, our dataset consists of the bulletins published from August 2015 to January 2019.}{}
%\Aron{if all results are based on Aug 15 -- Jan 19, then there is no reason to describe the preliminary scan (it may confuse the reader, who might think that this detail is important and influences the findings)} 
the website CVEDetails~\cite{Cve_Details}. In the following, we describe the details of our web scraper.

\subsection{Web Scraper}
\label{sub:web-based}

Considering that the mentioned websites do not provide JSON, XML or any exportable formats for the vulnerabilities and their details, it is essential to implement a web scraper. 
%\Aron{not sure if using BeautifulSoup and Selenium qualifies as ``from scratch,'' I'd just say that it is necessary to develop a crawler}
Our scraper is implemented in Python 3.6~\cite{Python_3_6_0}, with the help of its libraries called BeautifulSoup~\cite{BeautifulSoup} and Selenium Browser Automation~\cite{Selenium}. During the scraping, there is a possibility that the server blocks the client
%\change{a server can block the user}{the server blocks the client} 
if the server receives lots of consecutive requests from one particular IP address. %\Aron{``one particular IP address''?} 
We choose Selenium as it opens a web browser and navigates between the pages like a normal user does. Since it provides a small time gap between each consecutive request, it decreases the likelihood of blocking. 

%\Aron{how does this prevent sending too many requests? mimicking a human helps if the server tries to block automated requests, but this is not what the previous sentence describes}.

%By reason of having no JSON, XML or any exportable formats of all security bulletins, a scraper must have implemented from scratch. Hence, \textbf{Python 3.6} \cite{Python_3_6_0}, \textit{\textbf{Selenium Browser Automation}} \cite{Selenium},  \textbf{MongoDB} \cite{MongoDB}, \textbf{Studio 3T} \cite{Studio3T} and several Python libraries for implementing these technologies like \textbf{BeautifulSoup} \cite{BeautifulSoup}, \textbf{matplotlib}, \textbf{MongoClient} and \textbf{selenium} were used. For scraping, \textit{\textbf{selenium}} was chosen in purpose as too many requests at very short intervals had to be sent in a constant way. However, it would have caused to be blocked from the server if a server is configured to block such kind of requesting ways. For this reason, the requests had to be sent as if a user sends. Since Selenium opens a web browser, navigates between pages or types a text just like a user, it was our first choice to send our requests without getting blocked.

\subsubsection{Security Bulletins Dataset}
\label{subsub:features}

Figure~\ref{fig:dataset_entry}
%\change{The above snippet}{Figure~\ref{fig:dataset_entry}}
shows an example snippet 
%\change{}{snippet} 
from our JSON document, which illustrates
%\change{indicates}{illustrates} 
its general structure. At the first level, we use three keys for each security bulletin object. The first key is \textit{\textbf{timestamp}}, which shows the \textit{Unix timestamp} when a bulletin was
%\change{is}{was} 
published. 
%\change{ in the type of \textit{float}}{}\Aron{is this a Unix timestamp or something else? this could be mentioned}. 
The second key is \textbf{\textit{formatted\_date}}, which represents \textit{timestamp} in
%\change{that specifies}{which presents \textit{timestamp} in} 
a human-readable time format.
%\change{ of the field \textit{timestamp}}. 
The last key is \textbf{\textit{cves}}, which contains the details of all vulnerabilities and their patches that are published in a particular security bulletin. Under the \textit{cves} key, all vulnerabilities are grouped by their severity levels. There are four different severity levels in Android security bulletins: \textbf{low}, \textbf{moderate}, \textbf{high}, and \textbf{critical}. For example, if the severity level of a vulnerability is critical, then this vulnerability is placed under the \textit{critical} key. 
%\change{Since }{}
Android has its own security team, who
%\change{they}{who} 
define their own metrics for severity levels of  Android vulnerabilities~\cite{Andorid_Severity_Rankings}. As a result, these severity levels are different from commonly used severity levels such as version 2 and version 3 CVSS scores~\cite{NVD_Severity_Rankings}. In our security bulletin dataset, we use the Android security team's severity levels.

%Android has its own security team. The Android security team defines its own severity rankings [2] which are different from other publicly available datasets, like widely used version 2 and version 3 CVSS score [63]. In our dataset, we use severity rankings which is provided by Android security team

We also scrape the following information
%\change{columns}{information} 
from Android security bulletins: \textbf{CVE}, which is the ID of a vulnerability; \textbf{Type}; \textbf{Updated AOSP Versions}; \textbf{Date reported}; and \textbf{References}. Note that these fields
%\change{columns}{fields} 
might have different names on different security bulletins. For instance, the field %\change{column}{field} 
\textit{Updated AOSP Versions} has different names, such as \textit{Affected Versions} used in August 2015~\cite{Android_Security_Bulletin_2015_08} and \textit{Updated Versions} used in December 2015~\cite{Android_Security_Bulletin_2015_12}. 
To avoid confusion,
%\change{Therefore}{To avoid confusion}, 
we use the same
%\change{}{same} 
name \textbf{\textit{updated AOSP versions}} throughout this paper.

Similarly, the field
%\change{column}{field} 
\textit{References} also has
%\change{has also}{also has} 
more than one name. In early security bulletins, some of the following names have been used: ``\textit{Bug(s)}''~\cite{Android_Security_Bulletin_2016_01}, ``\textit{Bug(s) with AOSP Links}''~\cite{Android_Security_Bulletin_2015_12}, and ``\textit{Bugs with AOSP links}''~\cite{Android_Security_Bulletin_2016_02}. In the latest security bulletins, the term \textit{References} is used. As a result, we use \textbf{\textit{references}} throughout the paper for all above terms. Each reference is a link that navigates to a corresponding code repository. Note that there can be multiple references for a patched vulnerability that navigate to different 
%\Aron{what do you mean by branches?} 
branches of AOSP Git repositories. For instance, \texttt{CVE-2017-18159}, published in June 2018, has 9 different references~\cite{Android_Security_Bulletin_2018_06}. In the security bulletins, each of them is %\Aron{shown where?} 
listed with brackets in a separate row in the tables that consist of patched vulnerabilities. However, in early security bulletins, there are vulnerabilities whose references
%\change{that the references of each one of these vulnerabilities}{whose references} 
have 
%\Aron{what do you mean by ``separated by a row''? I don't think that readers will understand this, EXPLAINED BEFORE} 
%separated by 
multiple rows. Some of these different rows have even different severity levels and updated AOSP versions. For example, \texttt{CVE-2015-3873}, published in October 2015~\cite{Android_Security_Bulletin_2015_10}, has 14 different separate %\Aron{again, what is a ``row''? paper has not introduced columns or rows, so the reader will not know to what these refer, EXPLAINED BEFORE} 
rows. Except for the last row, all rows have the same severity level and the same updated AOSP versions and still belong to the same patched vulnerability. %This is the first issue we face. 
In such cases, where a vulnerability has more than one row, we check the severity level and the updated AOSP versions for each row.  If a row has the
same severity level and updated AOSP versions as previous rows, then we only add the new row's references into the existing references. Otherwise, we create a new entry for this particular row and consider this entry as a different patched vulnerability.

Another issue is that some vulnerabilities are patched on 
%\Aron{different compared to what? did you mean multiple?} 
multiple security bulletins. For instance, \texttt{CVE-2016-2059} is first %\change{firstly}{first} 
patched in September 2016, which is followed by another mentioning in October 2016~\cite{Android_Security_Bulletin_2016_09, Android_Security_Bulletin_2016_10}. Similarly, \texttt{CVE-2017-0391} is patched in both January 2017 \cite{Android_Security_Bulletin_2017_01} and June 2017. In such cases, we count them as different patched vulnerabilities in our dataset.

There are 80 vulnerabilities with these 
%\Aron{I'm not sure what the first issue was} 
two consistency issues. 36 of them have multiple rows, however, with the same severity levels and the same updated AOSP versions with multiple references. We count each of these 36 vulnerabilities only once. 
%are added into the original vulnerability references and not counted as a vulnerability more than once. 
In contrast, we count the remaining 44 vulnerabilities
%\change{rest, i.e., 44,}{remaining 44 vulnerabilities} 
multiple times, since they are on multiple
%\Aron{multiple?} 
different security bulletins or have multiple references with different severity levels and/or updated AOSP versions. %\textcolor{red}{THINK THIS PARAGRAPH MORE MORE TIME}

Furthermore, the attributes \textit{Date Reported} and \textit{Type} are not even used in some of the security bulletins, like September 2018 and July 2017~\cite{Android_Security_Bulletin_2018_09, Android_Security_Bulletin_2016_07}. In particular, for the attribute \textit{Date Reported}, Google stopped publishing it since June 2017~\cite{Android_Security_Bulletin_2017_06}. For this reason, in such cases, we leave the respective data fields empty. Hence, 
%\Aron{is ``if'' supposed to be here?} 
for certain analyses, we can only analyze the patched vulnerabilities that contain the necessary attributes.
%\Aron{what do you do with missing fields? some data imputation? leaving empty and dealing with them later in the analysis?} 

We add
%\change{For}{We add} 
the attributes that we scrape from CVEDetails 
%\change{, we add them} 
to the key named~\textbf{\textit{details}}. Under this key, there are vulnerability details such as \textit{public disclosure date}, \textit{CVSS score}, and \textit{products affected}, which are added to their corresponding key names. 

\subsubsection{Mining Code Repositories}
\label{subsub:mining_code_repositories}

Since RQ3, RQ4, and RQ5 are related to patching times of the vulnerabilities, %\Aron{not sure if reader will remember what these questions were, migth be worty adding a reminder that these pertain to patching times} 
we need to find the last fix-commit date of a vulnerability and the log of all changed lines on branches of each patched vulnerability. In such cases where Google does not provide references publicly, some vulnerabilities do not have any references. For the patched vulnerabilities that have references,  
there are three different reference types, which
link
to different code repositories. The first one is the AOSP Git repository~\cite{android_git_repo}, which is referenced by the
%\change{the case for}{referenced by the} 
majority of patched vulnerabilities. The second one is the Qualcomm Code Aurora page for Qualcomm-related vulnerabilities~\cite{Code_Aurora}. The third reference type is the Linux Patchwork page, which is used for kernel-related vulnerabilities~\cite{Linux_Patchwork}. Note that although there are vulnerabilities related to other vendors and third-parties (e.g., \textit{MediaTek} and \textit{libxml}), Google only published references to Qualcomm and Linux.

We scrape the %\Aron{Git branch?} 
\textbf{AOSP Git branches}, \textbf{directories of all changed files}, \textbf{the last commit dates}, and \textbf{the last commit IDs} from AOSP Git repositories. If a vulnerability has more than one reference, we scrape all of them. By collecting the above information, we can find which Android stack layer is affected by which vulnerability. We describe the details
%\change{The detail}{We describe the details} 
of our approach for finding the corresponding layer for each vulnerability 
%\change{is}{} 
in Appendix~\ref{app:layers}.
%Then, based on the classification method mentioned in \textcolor{red}{SECTION 5 -> APPENDIX}, we classify all the vulnerabilities.

\begin{figure}
\begin{lstlisting}[language=json, firstnumber=1, label=bulletin_vulnerability_json]
{
 "timestamp": 1538344800.0,
 "formatted_date" : ISODate("2018-10-01T00:00:00.000+0000"),
 "cves": {
  "critical": [
      {
        "id": "CVE-2018-9490",
        "type" : "EoP",
        "updated_aosp_versions" : "7.0, 7.1.1, 7.1.2, 8.0, 8.1, 9",
        "category" : "Framework",
        "references" : [
          {
            "name" : "A-111274046", 
            "link" : "https://android.googlesource.com/platform/external/chromium-libpac/+/948d..."
          },
          {
            "name" : "2", 
            "link" : "https://android.googlesource.com/platform/external/v8/+/a24543157..."
          }
        "details": {
          "cvss_score" : 0.0,
          "confidentiality_impact" : null,
          ...
        }
      }
      {
        ...
      }
    ]
  "high": [
    ...
   ]
  }
}
\end{lstlisting}
\caption{Example snippet from our security bulletin dataset}
\label{fig:dataset_entry}
\end{figure}
%\change{}{Example snippet from our security bulletin dataset.}

%\textcolor{red}{Scraping Code Repositories OR MINING CODE REPOSITORIES?}

For patched vulnerabilities that
%\change{}{that} 
originate from Qualcomm and Linux kernel, we only scrape the last fix-commit dates. However, not all vulnerabilities related to Qualcomm and 
%\Aron{kernel or Kernel?} 
Linux kernel have Qualcomm Code Aurora or Linux Patchwork pages, respectively. For example, \texttt{CVE-2018-10882}, published in January 2019~\cite{Android_Security_Bulletin_2019_01}, has a reference that points
%\change{navigates}{points} 
to a bug tracking tool called \textit{Bugzilla}. In such cases, we use the ticket closing time. 

After scraping the Android security bulletins, Git repositories and CVEDetails, we transfer all the retrieved data into a JavaScript Object Notation (JSON) document. JSON is an open-standard format that uses human-readable text to transmit data objects consisting of key-value pairs and array data types~\cite{JSON}. After formatting, we store the data %\change{it}{the data} 
in MongoDB~\cite{MongoDB}. For querying, displaying, and exporting our collected dataset in a JSON format, we use Studio 3T as a graphical user interface (GUI), which is a technology partner with MongoDB~\cite{Studio3T}. 
%For plotting the graphs, we use \textbf{matplotlib}~\cite{matplotlib}. 
In total, we collect 2,470 patched vulnerabilities from Android security bulletins that are published from August 2015 to January 2019 as well as their details from CVEDetails and Git repositories. Hereafter, we refer to
%\change{call}{refer to} 
our collected dataset that contains all Android security bulletins and CVEDetails as \textbf{security bulletins}.

%-------------------------------------------------------------------------------

%-------------------------------------------------------------------------------
\section{Results}
\label{sec:results}
In this section, we analyze the collected data with a focus on the posed
%\change{the}{of the} 
research questions. %and derive initial implications for consumer and technology policy. 
%\change{}{that we posed earlier} 
%\change{which are accompanied by discussions and their}{and discuss their} 

%Each of following subsection will belong to one research question.
%\textcolor{red}{DO WE MENTION THE CONSUMERS OR POLICIES?}

%\Aron{subsection title is very long, I suggest shortening, e.g., ``Evolution of Severity Distributions and Root Causes''}

\subsection{RQ1: Evolution of Severity Distributions and Root Causes}
%How have severity distribution and root causes of patched Android vulnerabilities evolved over time?}
%How have patched Android vulnerabilities evolved over time?}
%What is the overall picture on Android security patches?}
\subsubsection{Severity Levels}

%First, we investigate how severity levels of vulnerabilitieslook like over time and whether there exists some similaritiesamong them in different years. Note that each vulnerabilityhas a severity ranking defined by Android security team [2].Therefore, these severity levels may be different from com-monly used severity level methods, CVSS score V2 and CVSSscore V3.

We first investigate how severity levels of patched vulnerabilities evolve over time, and we examine the 
%\Aron{you mean similarities between years? not crystal clear to me}
similarities of the severity level trend between the years. Note that we use severity levels that are defined by the Android security team~\cite{Andorid_Severity_Rankings}, which are different from other used severity level calculations, such as V2 and V3 CVSS score.
%\change{ calculation}{}s.
%\textcolor{color}{Do you think it is good to also look at these differences? For example, how many of them are different, for which layer. Are Android severity levels more in compliance with V2 or V3?}
For instance, the Android security team classifies the severity level of a vulnerability as \textit{critical} if one of the following conditions is met: (i) Arbitrary code execution in the Trusted Execution Environment\footnote{Trusted Execution Environment is a component that is designed to be protected even from a hostile kernel.}, (ii) Remote arbitrary code execution in a privileged process, Bootloader, or the Trusted Computing Base\footnote{Trusted Computing Base is a part of the kernel, and it responsible for loading scripts into a kernel component.}, (iii) Remote permanent denial of service (device inoperability: completely permanent or requiring re-flashing the entire operating system). 
%the following criteria are counted a reason of ranking a vulnerability as critical:
%\begin{itemize}
%\item Arbitrary code execution in the TEE
%\item Remote arbitrary code execution in a privileged process, %Bootloader, or the TCB
%\item Remote permanent denial of service (device inoperability: completely permanent or requiring re-flashing the entire operating system)
%\end{itemize}
For the severity level of \textit{high}, the criteria are as follows: (i) Local secure Boot bypass, (ii) Remote arbitrary code execution in an unprivileged process, (iii) Local bypass of user interaction requirements on package installation or equivalent behavior.
%listed as follow:

%\begin{itemize}
%\item Local Secure Boot bypass
%\item Remote arbitrary code execution in an unprivileged process
%\item Local bypass of user interaction requirements on package %installation or equivalent behavior
%\end{itemize}

%\Aron{this is data cleaning, you might want to move it to the previous section}
Only one of the patched vulnerabilities does not have a severity level, %\change{i.e.,}{} 
\texttt{CVE-2016-2842}, which is published in August 2016 with the severity level of \textit{None*}~\cite{Android_Security_Bulletin_2016_08}. We 
%\Aron{do you exclude just for the severity level analysis or for all?} 
exclude this patched vulnerability
from our security bulletins dataset only for this analysis. Therefore, for our severity level analysis, we have 2,469 patched vulnerabilities. 
%In total, there are 2,469 patched vulnerabilities. The reason is that \texttt{CVE-2016-2842} that is published in August 2016 is excluded because it has the severity level of \textit{None*}~\cite{Android_Security_Bulletin_2016_08}.
Table~\ref{tab:annual} shows the annual number of patched vulnerabilities. In 2015, the number of patched vulnerabilities is for only 5 months. In subsequent years, the number of patched vulnerabilities is at least 7 times higher than 2015.

\begin{table}[h]
    \centering
    \small
    \begin{tabular}{|c|c|c|c|c|c|} \hline
        \textbf{Year} & \textbf{2015} & \textbf{2016} & \textbf{2017} & \textbf{2018} & \textbf{2019} \\ \hline
        No. of Vulnerabilities & 94 & 662 & 939 & 747 & 27 \\ \hline 
    \end{tabular}
    \caption{Annual number of patched vulnerabilities}
    \label{tab:annual}
\end{table}

\begin{figure}[h]
\centering
\includegraphics[width=\columnwidth]{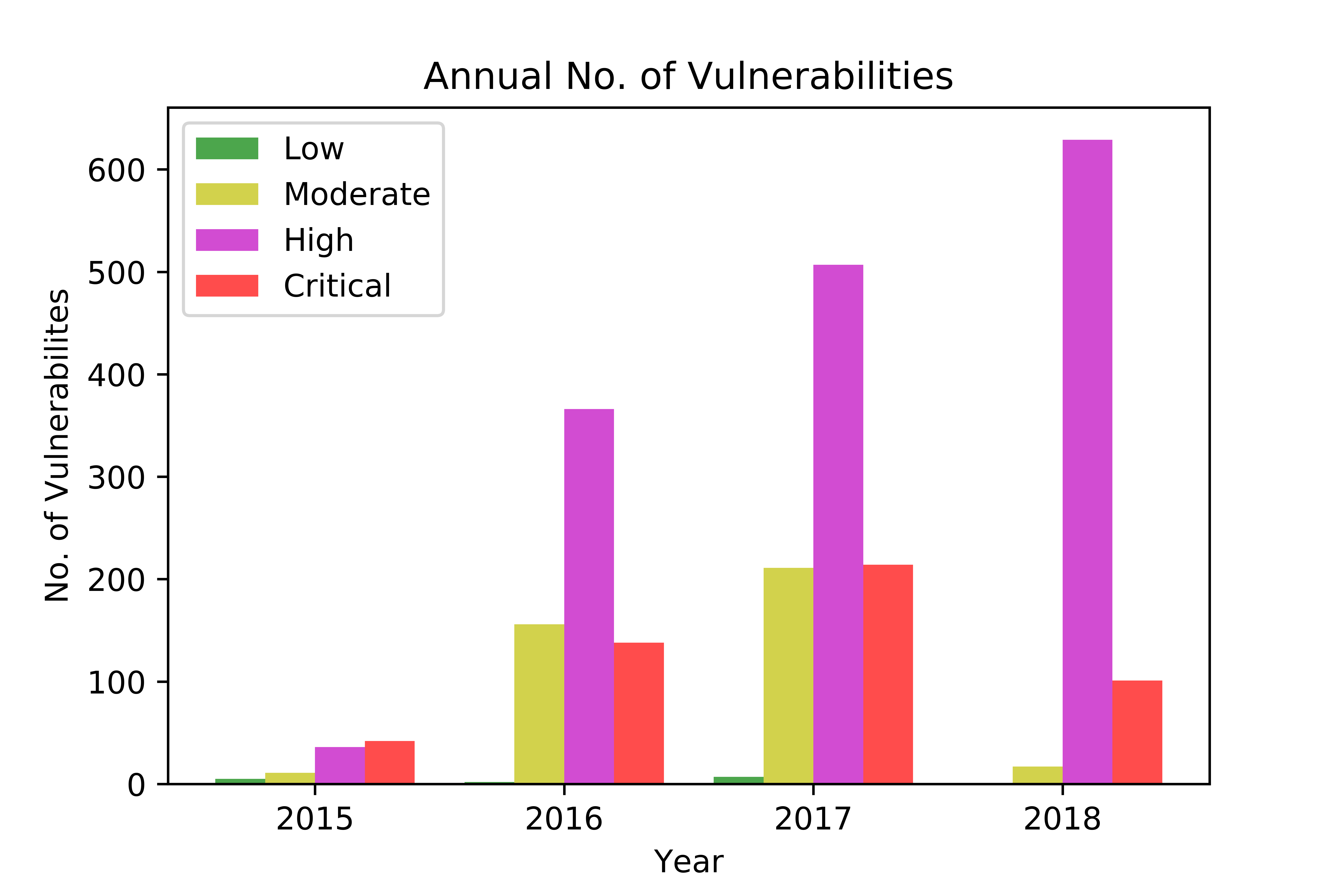}
\caption{Annual number of patched vulnerabilities by severity levels}
\label{fig:annual_severity}
\end{figure}

\begin{figure}[h]
\centering
\includegraphics[width=\columnwidth]{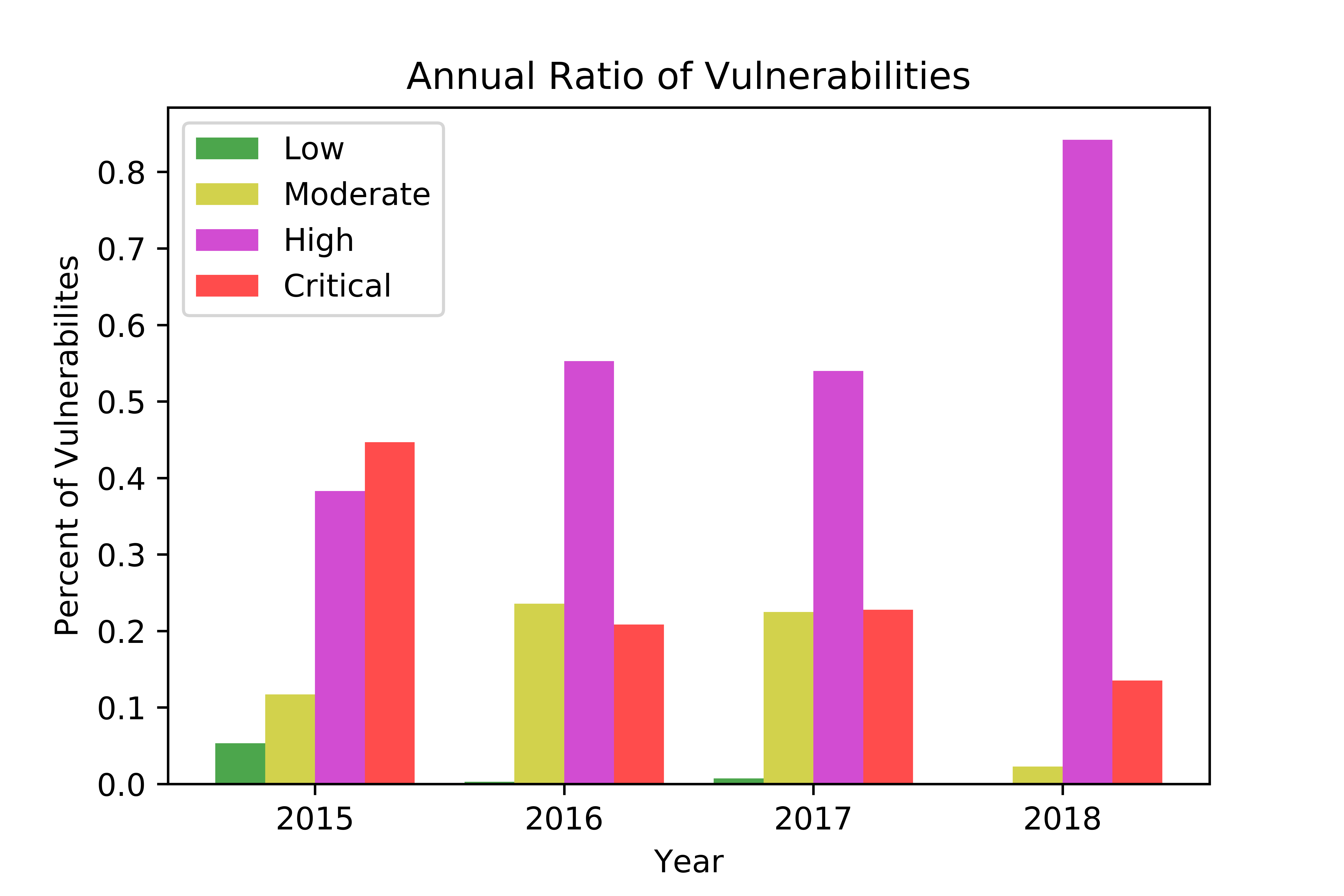}
\caption{Annual ratio of patched vulnerabilities by severity levels}
\label{fig:annual_severity_percent}
\end{figure}

Figure~\ref{fig:annual_severity} shows
%\change{represents}{shows} 
the annual number of patched vulnerabilities for each severity level from 2015 to 2018. Figure~\ref{fig:annual_severity_percent} also shows the annual percentage of each severity level (as a fraction of all vulnerabilities). Note that we do not consider the patched vulnerabilities in 2019 as we scrape only
%\change{only scrape}{scrape only} 
the January security bulletin. As we see in Figure~\ref{fig:annual_severity}, the number of  patched vulnerabilities with high severity level is increasing from 2015 to 2018. Furthermore, they are the majority among all patched vulnerabilities, except for 2015. In 2017, 507 of 939 patched vulnerabilities have a high severity level. In 2018, while the total number of patched vulnerabilities decreases from 939 to 747, the number of patched vulnerabilities with high severity level increases from 507 (54\%) to 629 (84\%). Similarly, the number of both moderate and critical severity levels rise slightly between 2016 and 2017. However, the number of moderate severity level patched vulnerabilities falls sharply between 2017 and 2018, from 211 to 17. 
%While the number of them is 211 of 939 in 2017, it decreases to 17 of 747 in 2018. 
In general, the number of
%\change{}{number of} 
low severity level patched vulnerabilities is 
%\change{are}{is} 
close to zero. In 2015 and 2016,
%\change{}{and 2016}, 
there are only 5 and 3
%\change{}{and 3} 
of them, respectively, and their number
%\change{. This number decreases to 3 in 2016 and}{, respectively, and their number} 
rises to only 7 in 2017. In 2018, we do not have \textit{any} (reported) patched vulnerabilities with low severity level.

%\textcolor{red}{Jens and Mehmet: Which figure is better? figure~\ref{fig:annual_severity} or Figure~\ref{fig:annual_severity_percent}} 
%Among these 2469 vulnerability patches, the following list expresses how many of them were patched in which year. 

%\begin{itemize}
%\item \textbf{2019}: 27
%\item \textbf{2018}: 747
%\item \textbf{2017}: 939
%\item \textbf{2016}: 662
%\item \textbf{2015}: 94
%\end{itemize}

%\begin{figure*}[t!]
%    \centering
%    \begin{subfloat}{width=0.4\textwidth}
%        \includegraphics[width=0.4\textwidth]{images/2015.png}
%    \end{subfloat}
%    \begin{subfloat}{width=0.4}
%        \includegraphics[width=0.4\textwidth]{images/2016.png}
%    \end{subfloat}
%    \begin{subfloat}[b]{width=0.4}
%        \includegraphics[width=0.4\textwidth]{images/2017.png}
%    \end{subfloat}
%     \begin{subfloat}[b]{width=0.4}
%        \includegraphics[width=0.4\textwidth]{images/2018.png}
%    \end{subfloat}
%    \begin{subfloat}[b]{width=0.4}
%        \includegraphics[width=0.4\textwidth]{images/2019.png}
%    \end{subfloat}
%    \caption{The number of patched vulnerabilities in monthly from the year 2015 to 2019 grouped by their severity rankings. Blue line: Critical, Orange Line: High, Grey Line: Moderate, Yellow Line: Low}
%    \label{fig:severity_rankings_monthly}
%\end{figure*}

%the number of patched vulnerabilities in subsequent years are at least 7 times higher than 2015, see Table~\ref{tab:annual}.

\begin{figure*}
\centering
\includegraphics[width=\textwidth]{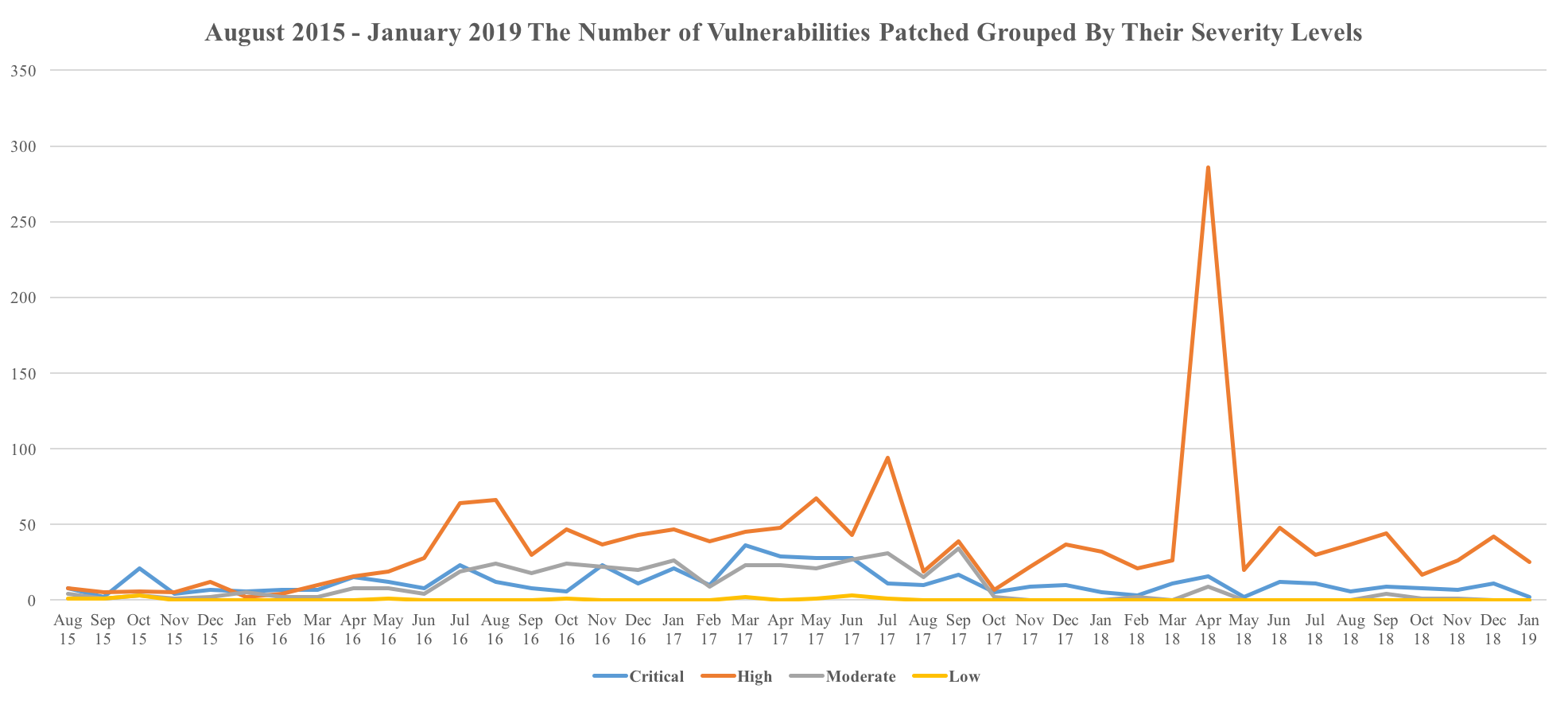}
\caption{Number of monthly patched vulnerabilities from August 2015 to January 2019 grouped by their severity levels}
\label{fig:severity_rankings_monthly}
\end{figure*}

Figure~\ref{fig:severity_rankings_monthly} depicts the number of %\change{monthly}{} 
patched vulnerabilities for each month
%\change{}{for each month} 
from August 2015 to January 2019 grouped by severity levels.
In this figure, we observe that starting from March 2016, the high severity level patched vulnerabilities are always the most frequent. In April 2018, we see the most pronounced peak of high severity level patched vulnerabilities; to a lesser degree there is also a peak in July 2017. In both these security bulletins, there is a separate section that
%\change{}{that} 
mentions cumulative updates of Qualcomm closed-source components. For example, in April 2018, there are 286 patched vulnerabilities that belong to the cumulative update~\cite{Android_Security_Bulletin_2018_04}. 240 out of 286 are high severity patched vulnerabilities.%, which explain why their number is
%\Aron{the number of them are}{their number is}
%so high in April 2018. 
Similarly, there are 94 patched vulnerabilities that belong 
%\change{belong}{that belong} 
to the cumulative update of Qualcomm closed-source components in July 2017~\cite{Android_Security_Bulletin_2017_07}. Table~\ref{tab:mean_severity_rankings_yearly} %\Aron{what is $X$? it has not been defined (do yo need this symbol? same goes for $\sigma$} 
and Table~\ref{tab:stdev_severity_rankings_yearly} show the mean and standard deviation values, respectively,
%\change{}{, respectively,} 
of the monthly number of patched vulnerabilities for each severity level annually.
%\change{, respectively}{}. 

\begin{table}[h]
    \centering
    \begin{tabular}{|c|c|c|c|c|} \hline
        \textbf{Severity Level} & \textbf{2015} & \textbf{2016} & \textbf{2017} & \textbf{2018} \\ \hline
        Critical & 3.5 & 11.5 & 17.83 & 8.41 \\ \hline
        High & 3 & 30.5 & 42.25 & 52.41 \\ \hline
        Moderate & 0.91 & 13 & 17,58 & 1,41 \\ \hline
        Low & 0.41 & 0.25 & 0.58 & 0 \\ \hline
    \end{tabular}
    \caption{Mean of monthly number of patched vulnerabilities for each  severity level annually}
    \label{tab:mean_severity_rankings_yearly}
\end{table}

\begin{table}[h]
    \centering
    \begin{tabular}{|c|c|c|c|c|} \hline
        \textbf{Severity Levels} & \textbf{2015} & \textbf{2016} & \textbf{2017} & \textbf{2018} \\ \hline
        Critical & 6.23 & 6.05 & 10.20 & 4.05 \\ \hline
        High & 4.11 & 21.61 & 22.55 & 74.22 \\ \hline
        Moderate & 1.37 & 8.89 & 12.13 & 2.67 \\ \hline
        Low & 0.90 & 0.45 & 0.99 & 0 \\ \hline
    \end{tabular}
    \caption{Standard deviation of monthly number of patched vulnerabilities for each  severity level annually}
    \label{tab:stdev_severity_rankings_yearly}
\end{table}

%\change{In order t}{T}
To check whether each of moderate, high, and critical severity levels has statistically significant differences in 
%\change{terms of}{} 
their monthly means between years,
%\change{for each year}{between years}, 
we perform several different Analysis of Variance (ANOVA) tests. Note that we exclude low severity patched vulnerabilities as we do not have enough samples to justify a meaningful result. Our null hypothesis for all of our tests is the same: For each of these three severity levels, the %\Aron{observed means are obviously different, so this must study the mean of their distributions} 
population means of their
%\change{}{their} 
monthly patched vulnerabilities are the same in every year,
%\change{as each other for each year}{in every year}, 
given confidence level $\alpha = 0.05$. For the first ANOVA test, we consider the absolute number of patched vulnerabilities in each month for a severity level and compare between
%\change{comparing}{compare} \change{them with}{between} 
different years. Table~\ref{tab:ANOVA_monthly_absolute_all} depicts the results of our ANOVA analysis. F-values are relatively high and $p$-values are lower than our confidence level, so
%\change{}{so} 
we can reject the null hypothesis. This means that the mean number of patched vulnerabilities in each year is different from other years for moderate, high, and critical severity levels.

\begin{table}[]
    \centering
    \begin{tabular}{|c|c|c|} \hline
         \textbf{Severity Level} & \textbf{F-value} & \textbf{$p$-value}  \\ \hline
         Critical & 8.85566 & 0.00001048 \\ \hline
         High & 3.35922 & 0.0270576 \\ \hline
         Moderate & 14.2759 & 0.0000012 \\ \hline
    \end{tabular}
    \caption{F and $p$-values of ANOVA test performed on absolute values for different severity levels of all patched vulnerabilities}
    \label{tab:ANOVA_monthly_absolute_all}
\end{table}

%The second ANOVA test is taking the percentages of the values rather than taking the absolute values. In other words, the numbers of monthly patched vulnerabilities are the percentages in a corresponding year. Table~\ref{tab:ANOVA_monthly_percentage_all} shows the results.

We also perform an ANOVA test on the monthly percentage of each severity level (as a fraction of all vulnerabilities) whose results are shown in Table~\ref{tab:ANOVA_monthly_percentage_all}. Here, we cannot reject our null hypothesis.
%\change{}{(as a fraction of all vulnerabilities)}, 
%\change{which is}{whose results are}
%Since the F-values are lower than the values of Table~\ref{tab:ANOVA_monthly_absolute_all} and $p$-value is almost equal to 1, we cannot reject our null hypothesis. 
This means that if we consider
%\change{take}{consider} 
the percentages of patched vulnerabilities of a severity level (as a fraction of all vulnerabilities) instead of
%\change{grouped by their severity levels rather than}{of a severity level (as a fraction of all vulnerabilities) instead of} 
their absolute values, the means of each severity level are uniform over the years.
%\change{in different years are equal}{are uniform over the years}.

\begin{table}[]
    \centering
    \begin{tabular}{|c|c|c|} \hline
         \textbf{Severity Level} & \textbf{F-value} & \textbf{$p$-value}  \\ \hline
         Critical & 1.36501 & 0.99999999 \\ \hline
         High & 4.65718 & 0.99999999 \\ \hline
         Moderate & 2.02166 & 0.99999999 \\ \hline
    \end{tabular}
    \caption{F and $p$-values of ANOVA test performed on percentage values for different severity levels of all patched vulnerabilities}
    \label{tab:ANOVA_monthly_percentage_all}
\end{table}

%we cannot say the same as the F-values are lower than the first table and $p$-value is almost equal to 1. This means that if we take the percentages of the values, rather than their absolute values, the means are most likely the same.

To examine whether Qualcomm-related vulnerabilities have a significant effect on the mean numbers, we also perform ANOVA tests by excluding these vulnerabilities.
%\change{By excluding the Qualcomm-related patched vulnerabilities, we also perform the ANOVA tests in order to examine whether they have a huge effect on changing the mean of the samples.}{To examine whether Qualcomm-related vulnerabilities have a significant effect on the mean numbers, we also perform ANOVA tests by excluding these vulnerabilities.}
Based on Figure~\ref{fig:severity_rankings_monthly}, there are peak points that Qualcomm patched vulnerabilities cause. Table~\ref{tab:ANOVA_monthly_absolute_without_qualcomm} and Table~\ref{tab:ANOVA_monthly_percentage_without_qualcomm} show %\change{demonstrate}{show} 
the ANOVA test results for absolute and percentage values, respectively. Here, we observe results that are similar to those including the Qualcomm vulnerabilities.
%\change{a similar behavior}{results that are similar to those that include the Qualcomm vulnerabilities}. 
For absolute values, the means are not equal, while for percentage values, the population means are equal in different years.
%Although the F-values are high and $p$-values are lower than our confidence level in Table~\ref{tab:ANOVA_monthly_absolute_without_qualcomm}, it is the opposite for Table~\ref{tab:ANOVA_monthly_percentage_without_qualcomm}.
In sum, our
%\change{Our above}{In sum, our} 
analysis shows that even though the average number of patched vulnerabilities for each severity level changes annually, their 
%\change{but the}{their} 
average percentage (as a fraction of all vulnerabilities)
%\change{of them}{(as a fraction of all vulnerabilities)} 
is the same over years. Therefore, we can expect distributions with the same average percentages
%\change{behavior}{distributions} 
in the future. 
%\change{Most of patched vulnerabilities have high severity level.}{} 

%\textcolor{red}{SOME DISCUSSIONS?}

\begin{table}[]
    \centering
    \begin{tabular}{|c|c|c|} \hline
         \textbf{Severity Level} & \textbf{F-value} & \textbf{$p$-value}  \\ \hline
         Critical & 6.60945 & 0.00087576 \\ \hline
         High & 11.28015 & 0.00087576 \\ \hline
         Moderate & 14.14615 & 0.00087576 \\ \hline
    \end{tabular}
    \caption{F and $p$-values of ANOVA test performed on absolute values for different severity levels by excluding patched Qualcomm vulnerabilities from the security bulletin dataset}
    \label{tab:ANOVA_monthly_absolute_without_qualcomm}
\end{table}

\begin{table}[]
    \centering
    \begin{tabular}{|c|c|c|} \hline
         \textbf{Severity Level} & \textbf{F-value} & \textbf{$p$-value}  \\ \hline
         Critical & 4.51910 & 0.99999999 \\ \hline
         High & 4.51910 & 0.99999999 \\ \hline
         Moderate & 3.99611 & 0.99999999 \\ \hline
    \end{tabular}
    \caption{F and $p$-values of ANOVA test performed on percentage values for different severity levels by excluding patched Qualcomm vulnerabilities from the security bulletin dataset}
    \label{tab:ANOVA_monthly_percentage_without_qualcomm}
\end{table}

\subsubsection{Common Weakness Enumerations}
\label{subsub:CWE}
 
CWE is a formal list which describes a common language to classify software security weaknesses. \texttt{Buffer Overflows}, \texttt{Structure and Validity Problems}, and \texttt{Authentication Errors} are some examples of enumeration categories~\cite{CWE}. We analyze these enumerations because they specify what causes vulnerabilities. Therefore, we can determine
%\change{demonstrate}{determine} 
what are the common software weaknesses among the patched vulnerabilities. Note that CWE information is
%\change{they are}{CWE information is} 
not indicated on security bulletins. Thus, we use the field \textit{CWE ID} and its details, which we scrape from the web page CVEDetails. 

%CWE is a formal list of software weakness types. It describes a common language to classify software security weaknesses.

%In order to analyze the CWE types of Android patched vulnerabilities, we need to look at the details of them that we scraped from CVEDetails. 
%Here the ids, the categories and the descriptions of CWE are indicated.

Among the 2,470 patched vulnerabilities, 109 
%\change{of them}{} 
do not have any details listed
%\change{indicated}{listed} 
on the CVEDetails web page.
%\change{ web page}. 
In addition, 58 vulnerabilities
%\change{of them}{vulnerabilities} 
have details but do not have any CWE. After excluding these, we continue our analysis with the remaining 2,303 patched vulnerabilities.

Among the common weakness enumeration categories, there is a parent-child hierarchy
%\change{ between each of them}{}
\cite{CWE_Hierachy}. For instance, \texttt{CWE-306: Missing Authentication for Critical Function} is a child of \texttt{CWE-285: Improper Authorization}. Furthermore, this relation can be one-to-many. In other words, there are
%\change{might be}{are} 
enumerations that are a child of more than one other enumeration. For instance, \texttt{CWE-358: Improperly Implemented Security Check for Standard} is a child of both \texttt{CWE-693: Protection Mechanism Failure} and \texttt{CWE-254: Security Features}. Moreover, \texttt{CWE-693} is a child of \texttt{CWE-254: Security Features}. Hence, \texttt{CWE-254} should also be considered when a vulnerability belongs to the CWE category of 
%\change{has}{is} 
\texttt{CWE-358}. 

%As an example, let's consider CWE-306, i.e., Missing Authentication for Critical Function. CWE-306 is child of CWE-285, i.e., Improper Authorization. CWE-285 is child of CWE-284, i.e., Improper Access Control. Contrary to this example, this parent-child relation is not one-to-one. In other words, there are some CWE categories that are child of more than one categories. For example, CWE-358 (Improperly Implemented Security Check for Standard) is child of both CWE-693 (Protection Mechanism Failure) and CWE-254 (Security Features). Moreover, CWE-693 is child of CWE-254. Hence, in our analysis, CWE-693 should also be considered when a vulnerability has the category of CWE-358. 

For a concrete example, consider \texttt{CVE-2017-0757}, which was
%\change{}{which was} 
published in September 2017~\cite{Android_Security_Bulletin_2017_09}. According to the CVEDetails web page, this vulnerability
%\change{it}{this vulnerability} 
has the enumeration of \texttt{CWE-284: Access Control (Authorization) Issues}. Since \texttt{CWE-284} has three different parent enumerations, which are \texttt{CWE-693: Protection Mechanism Failure}, \texttt{CWE-264: Permissions, Privileges and Access Control}, and \texttt{CWE-664: Improper Control of a Resource Through its Lifetime}, we also consider these three parent enumerations. 

%\change{Particularly, \texttt{CVE-2017-0757} belongs not only \texttt{CWE-284}, but also its parents and parents of its parents.}{}
%\textcolor{red}{LAST SENTENCE?}

%To give a concrete example on an Andriod vulnerability, let's consider CVE-2017-0757. According to CVEDetails, it has the the category of CWE-284, i.e., Access Control (Authorization) Issues. Since CWE-284 has three different parent category which are CWE-693 (Protection Mechanism Failure), CWE-264 (Permissions, Privileges and Access Control) and CWE-664  (Improper Control of a Resource Through its Lifetime), we count them all for CVE-2017-0757. 

\begin{figure}[h]
\centering
\includegraphics[width=\columnwidth]{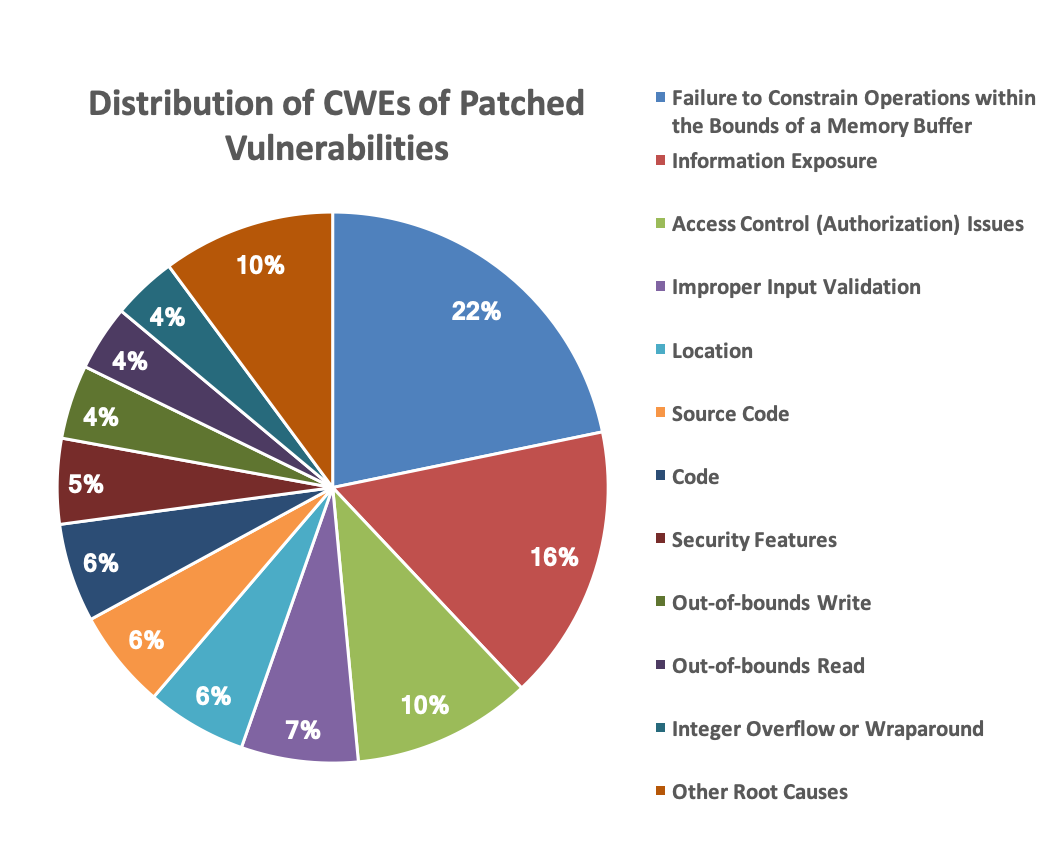}
\caption{Distribution of CWEs of Patched Vulnerabilities}
\label{fig:overall_distribution_of_cwe_categories}
\end{figure}

\begin{figure}[h]
\centering
\includegraphics[width=\columnwidth]{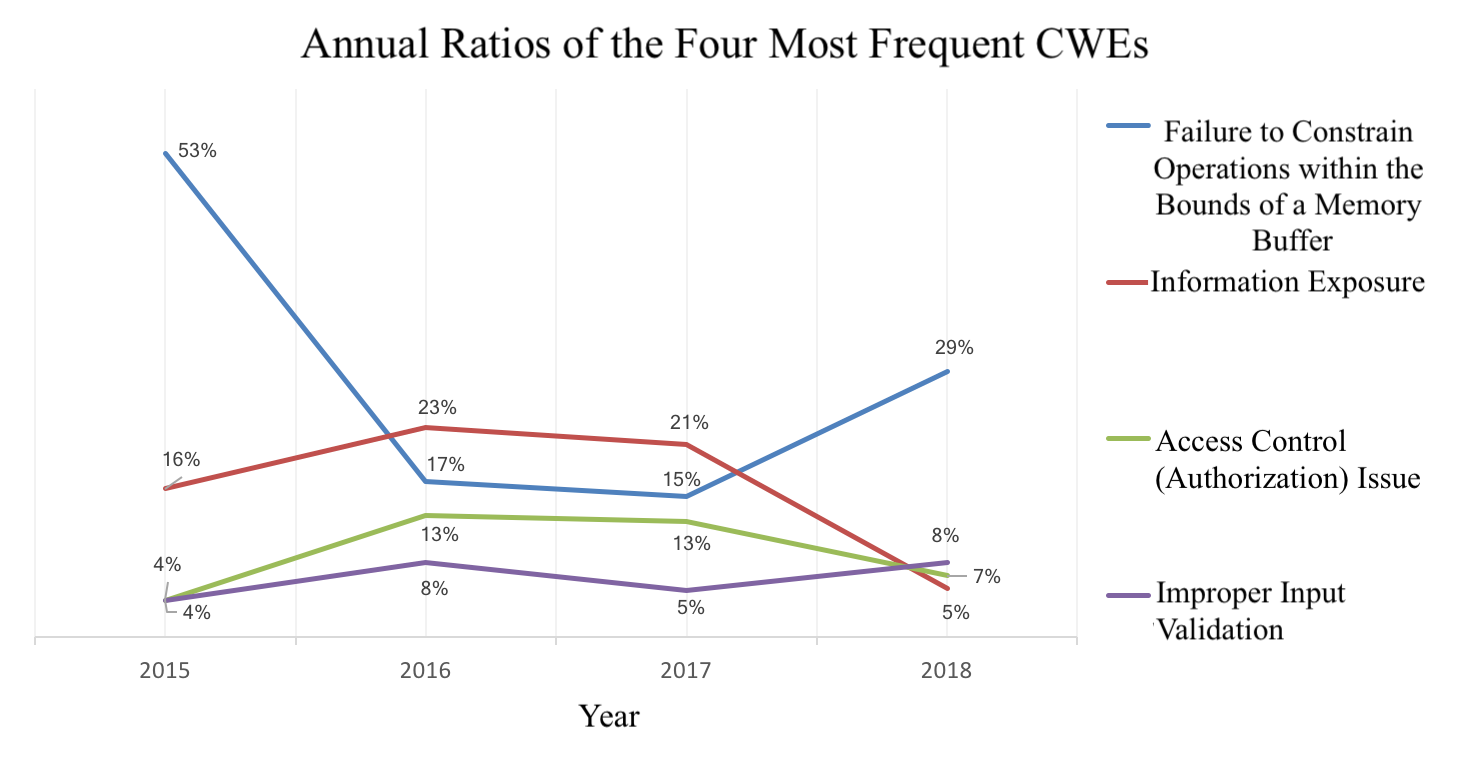}
\caption{Annual ratios of the four most frequent CWEs}
\label{fig:the_delta_four_most_cwe_types}
\end{figure}

%\change{Given this explanation, t}{T}
The overall distribution of CWE categories among the Android patched vulnerabilities and the annual percentage of the four most frequent CWE categories are plotted in Figure~\ref{fig:overall_distribution_of_cwe_categories} and Figure~\ref{fig:the_delta_four_most_cwe_types}, respectively.

Based on Figure~\ref{fig:overall_distribution_of_cwe_categories}, the four most frequent CWE categories make up more than 50\% of all the vulnerabilities patched. The reason for patching mostly these CWE categories might be that the developers---especially those who lack coding experience or familiarity with language features---tend to introduce such weaknesses quite often.  

Figure~\ref{fig:the_delta_four_most_cwe_types} shows the annual percentages of the four most frequently patched CWE categories. %based on Figure~\ref{fig:overall_distribution_of_cwe_categories}. 
According to Figure~\ref{fig:the_delta_four_most_cwe_types}, in contrast with the stability of the severity level ratios, there is more variation in the occurrence of the four most frequent CWE categories. For instance, although \texttt{CWE 119: Failure to Constrain Operations within the Bounds of a Memory Buffer} is the most frequently patched category in 2015 and 2018, it is only the second most frequent one in 2016 and 2017.

%The third most patched CWE type is CWE 284 (Access Control- Authorization Issues). If any of these following protection mechanisms is not applied properly, an attacker can gain privileges or read sensitive information~\cite{CWE_284}. Although it decreases to 4\% in 2018, it has no critical changes in the years 2015, 2016 and 2017, see Figure \ref{fig:the_delta_four_most_cwe_types}. However, this CWE type is always one of the critical CWE types that should been considered seriously.

%The fourth most patched CWE type is CWE 20 (Improper Input Validation). If an attacker can craft the input in a form because of not validating the input properly and alter the control flow or arbitrary control of a resource~\cite{CWE_20}. Based on Figure \ref{fig:overall_distribution_of_cwe_categories}, 7\% of the total patched vulnerabilities falls into this type. According to Figure~\ref{fig:the_delta_four_most_cwe_types}, this CWE type does not fluctuate too much compared to the other three CWE types over the years. 
%even though it reaches to 10\% in 2018.

%Potential reasons for these unknown CVE IDs
%\begin{itemize}
%    \item They are released recently
%    \item We probably see less number of unknown CVEs for 2018 over time
%    \item The number of unknown CVEs in each month is not related to the number of CVEs on that month on Android security bulletin
%    \item Most of these CVEs have severity level of high.
%\end{itemize}

\subsection{RQ2: Security Support Duration}
%What is the effect of ending support for old versions on consumers' security?}
\label{sub:eol}
The Android Open Source Project (AOSP) is a ready to be released version of Android. The source code of AOSP can be customized and adapted by original equipment manufacturers (OEMs) to run on their devices~\cite{All_AOSP_Versions}. 
In this subsection, we investigate the number of patched vulnerabilities that affect each AOSP versions. 
Android has 28 different API levels and 63 different AOSP versions.
In our security bulletins dataset, there exist 15 different AOSP versions. Note that Android was released in 2008 and the first security bulletin was published in August 2015. Thus, some AOSP versions have reached their end of life before the first bulletin, so
%\change{, and do not receive security updates anymore. Therefore,}{before the first bulletin, so} 
we only see a limited number of AOSP versions in the Android security bulletins. Further, we exclude AOSP version 6.1 from
%\change{in}{from} 
our analysis because
%\change{. The reason is that}{ because} 
we only see two patched vulnerabilities for this version: \texttt{CVE-2016-2496} and \texttt{CVE-2016-3843}, which were
%\change{are}{were} 
published in June 2016 and August 2016, respectively~\cite{Android_Security_Bulletin_2016_06, Android_Security_Bulletin_2016_08}. Therefore, in our analysis, we consider %\change{have}{consider} 
only 14 different AOSP versions. 

%In this subsection, we investigate the number of vulnerability patches that affect each AOSP versions that are published in Android security bulletins. AOSP is ready to be released version of Android. The source code of AOSP can be customized and adapted by original equipment manufacturers (OEMs) to run on their devices~\cite{All_AOSP_Versions}. 
%Android has 28 different API levels and 63 different AOSP versions.
%In our security bulletins dataset, there exists 15 different AOSP versions. Note that Android was released in 2008 and the first security bulletin was published in August 2015. Thus, it is obvious that some AOSP versions have reached to their end of life, and do not receive security updates anymore. Therefore, we only see limited number of AOSP versions in Android security bulletins. Further, in our analysis, we exclude AOSP version 6.1. The reason is that we only see two CVEs for this version: CVE-2016-2496 and CVE-2016-3843. Therefore, in our analysis, we have only 14 different AOSP versions.

%\change{Before starting to analyze different AOSP versions, we observe the following terms for them on e}{E}
Early Android security bulletins do not specify AOSP versions but instead use the terms
%\change{:}{do not specify AOSP versions but instead use the terms} 
\textit{5.1 and below}, \textit{6.0 and below}, \textit{5.0 and above}, and \textit{6.0 and above}. As a result, we first
%\change{}{first} 
need to map
%\change{find suitable AOSP versions for both}{map} 
the terms \textit{below} and \textit{above} to AOSP versions.
%\change{}{to AOSP versions}. 
%\change{Here, w}{W}
We use a conservative approach which is 
%\change{described}{} 
as follows: 
%\change{Since there is version number followed by either \textit{below} or \textit{above}, we can formulate this expression as follows: \texttt{$\mathcal{X}$ and below/above}.}{} 
For the term \textit{X and below}, we consider version $\mathcal{X}$ as well as versions that were introduced earlier than version $\mathcal{X}$. For the term \textit{$\mathcal{X}$ and above}, on the other hand, we consider version $\mathcal{X}$ as well as versions that were introduced later than version $\mathcal{X}$ but earlier than the security bulletin that we are studying.
%\change{are higher than the version $\mathcal{X}$ and have been introduced before than the publication date of a security bulletin that contains this patched vulnerability}{were introduced later than version $\mathcal{X}$ but earlier than the security bulletin that we are studying} 
For example, we replace the expressions \textit{5.1 and below} and \textit{5.0 and above} with \{\textit{4.4, 4.4.4, 5.0, 5.0.2, 5.1}\} and \{\textit{5.0, 5.0.2, 5.1}\}, respectively.

\begin{figure}[h]
\centering
\includegraphics[width=\columnwidth]{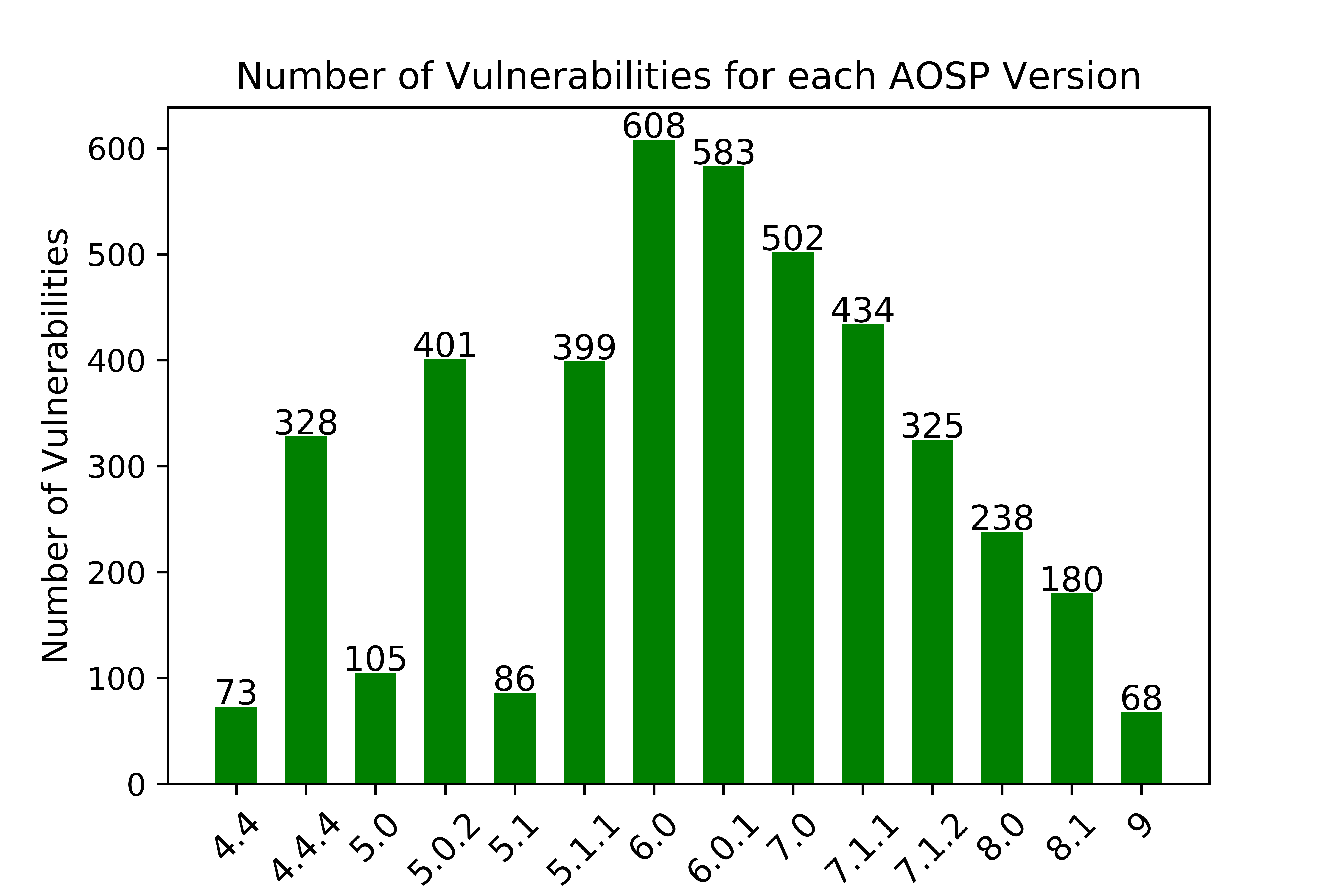}
\caption{The number of patched vulnerabilities for each AOSP version}
\label{fig:number_of_vulnerabilities_affect_each_aops_versions}
\end{figure}

As we mentioned earlier, there are 2,470 patched vulnerabilities. Among these, 889 
%\change{of them}{} 
have an entry in the field for updated AOSP versions in the Android security bulletins. 
Figure~\ref{fig:number_of_vulnerabilities_affect_each_aops_versions} shows the number of vulnerabilities for each AOSP version from August 2015 to January 2019. 

%In the following, we investigate the common patched vulnerabilities among different AOSP versions and %\Aron{inconsistencies between what?} 
%those CVE IDs in Android security bulletins that have unknown CVE ID in CVEDetails.
%inconsistencies on Android security bulletins and CVEDetails.

\subsubsection{Release Date and Update Duration}

In this subsection, we conduct pairwise comparisons for consecutive AOSP versions focusing on the gap in release time between the two versions vs. the gap in the time when the final patch for a particular version was released. If all versions are treated equally in terms of support, we would expect that these two gaps are identical. Also, we would expect that the data is similar over time for different pairs of versions.

\begin{figure}[h]
\centering
\includegraphics[width=\columnwidth]{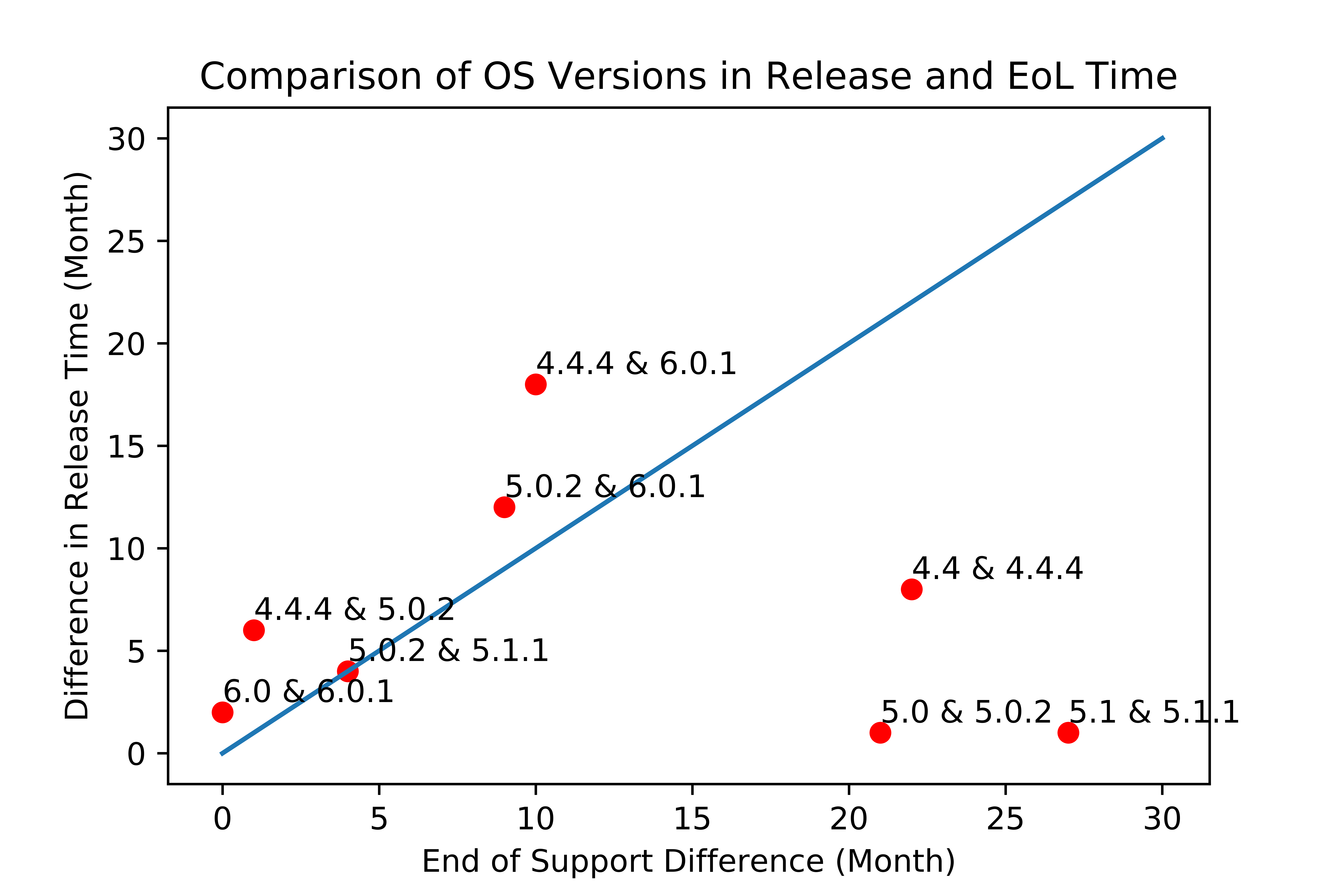}
\caption{Pairwise comparison of time differences in release time and end-of-life in Android security bulletins. Here, the blue line shows that the difference in release time and end-of-life is equal for two versions}
\label{fig:Release_End}
\end{figure}

In our analysis, we exclude the versions 7.0, 7.1.1, 7.1.2, 8.0, 8.1, and 9 because these versions are still receiving updates at the time of our data collection
%\change{to the time that we have collected our dataset}{at the time of our data collection} 
(January 2019). The result of our analysis is presented in Figure~\ref{fig:Release_End}.
%\Aron{this is very unclear to me, which version is compared to which (e.g., for versions 4.4 and 4.4.4, is the difference from 4.4 to 4.4.4 or vice versa?) for release time, it is clear that later version always come later; however, for EOL time, this is not necessarily true}. 
In this figure, the horizontal axis
%\change{x-axis}{horizontal axis} 
shows the time difference of end-of-life (EOL) between two versions in months. Note that we calculate the EOL based on the last time that we see an update for an AOSP 
version on Android security bulletins. The vertical axis
%\change{y-axis}{vertical axis} 
represents the time difference in the release time of two versions in months. Each point indicates these two differences for two versions. The blue line represents the point where the time difference of EOL and release time for the two versions are equal.

%In this subsection, we compare the difference in release time of different versions and its relation to the duration that they received security updates. In our analysis, we exclude versions 7.0, 7.1.1, 7.1.2, 8.0, 8.1, and 9 because these versions are still receiving updates to the time that we have collected our dataset (January 2019). The result of our analysis is represented in Figure~\ref{fig:Release_End}. In this figure, the x-axis shows the time difference in the end-of-life between two versions in month. Note that we calculate the end-of-life based on the last time that we see an update for an AOSP version in Android security bulletin. The y-axis represents the time difference in the releasing time of two versions in month. Each point in this figure indicates these two differences for two versions and the blue line represents the point where these two time differences are equal.

According to Figure~\ref{fig:Release_End}, the pair consisting of version 5.0.2 and version 5.1.1 is the only one located on the blue line. Version 5.1.1 was introduced 4 months after version 5.0.2. Also, version 5.0.2 stopped receiving updates 4 months sooner than version 5.1.1. 

%According to Figure~\ref{fig:Release_End}, version \textit{5.0.2} and version \textit{5.1.1} are the only versions that are on the blue line. In other words, version \textit{5.1.1} was introduced 4 months after version \textit{5.0.2} and version \textit{5.0.2} stopped receiving updates 4 months sooner than version \textit{5.1.1}. 

Points that are under the blue line are those pairs of versions where the newer version continues receiving updates comparatively longer than the difference in their release dates would indicate.

There are three points in that region. We describe each of them in detail in the following. % WHY ONLY THREE POINTS?

%Points that are under the blue line are those versions that the older one continues receiving updates longer than the time difference in their release. There are three points in that region. We describe each of them in detail in the following.

\begin{itemize}
    \item The difference in introduction date of version 4.4 and 4.4.4 is about 8 months. However, version 4.4 stopped receiving updates 22 months sooner.
    
    \item Version 5.0.2 was introduced one month after 5.0. Version 5.0.2 continued receiving updates 21 months after the last update date of version 5.0.
    
    \item Version 5.1 was introduced one month before version 5.1.1. The former stopped receiving updates 27 months sooner than version 5.1.1. %[Furthermore, version 5.1 was introduced 4 and 3 months, respectively, after version 5.0 and 5.0.2. However, 5.1 stopped receiving updates relatively early  in comparison to these two versions.] 
\end{itemize}

%\begin{enumerate}
 %   \item The different in introduction of version 4.4 and version 4.4.4 is near 8 months. However, version 4.4 stopped receiving updates 22 months sooner.
    
  %  \item Version 5.0.2 was introduced one month after 5.0. Version 5.0.2 continued receiving updates 21 months after the last update of version 5.0.
    
   % \item Version 5.1.1 was introduced one month after version 5.1, while it stopped receiving updates 27 months sooner than version 5.1.1. Furthermore, version 5.1 was introduced 4 and 3 months after version 5.0 and 5.0.2, respectively. However, it stopped receiving updates sooner than both of them. 
%\end{enumerate}

%COMMENTED OUT -- WHERE IS THE DATA TO SHOW THIS: Based on the detailed observations above, we only see comparatively short periods of patch updates for older versions among subversion of AOSP versions, like 4.4. to 4.4.4. However, we do not see such behavior for AOSP versions of 6 or higher.

The points above the blue line are those pairs of versions where the gap between release times is larger than the gap between EOL times. We have 4 pairs of such situations that are also plotted in Figure~\ref{fig:Release_End}.

%The points above the blue line are those versions that the older version stopped receiving updates before the newer version, but it receives updates for longer period of time by knowing their release time. For that region, we have 4 points that we explain them in the following.
\begin{itemize}
    \item Version 6.0 and version 6.0.1 stopped receiving updates at the same time while 6.0.1 was introduced 2 months after version 6.0.
    
    \item Version 5.0.2 was introduced 6 months after version 4.4.4. However, version 4.4.4 stopped receiving updates just one month sooner than version 5.0.2.
    
    \item Version 4.4.4 was introduced 18 months sooner than version 6.0.1. But, it stopped receiving updates just 10 months before version 6.0.1.
    
    \item Version 6.0.1 was introduced one year after version 5.0.2. But, version 6.0.1 continues receiving updates only 9 months after version 5.0.2.
    %\item All other versions that we take into account, versions 7.0, 7.1.1, 7.1.2, 8.0, 8.1, and 9, are still receiving updates to the time that we have collected our dataset (January 2019).
    
    %\item We see the first update for version 5.1.1 on November of 2015 (6.0 and below) which was released on April 2015 which is a long time. Furthermore, it is interesting to see that the first update for version 6.0 is on November 2015 which is also released at November 2015.
\end{itemize}

%\begin{enumerate}
 %   \item Version \textit{6.0} and version \textit{6.0.1} stopped receiving updates at the same time while \textit{6.0.1} was introduced 2 months after version \textit{6.0}.
    
  %  \item Version \textit{5.0.2} was introduced 6 months after version \textit{4.4.4}. However, version \textit{4.4.4} stopped receiving updates one month sooner than version \textit{5.0.2}.
    
  %  \item Version \textit{4.4.4} was introduced 18 months sooner than version \textit{6.0.1}. But, it stopped receiving updates 10 months before version \textit{6.0.1}.
    
  %  \item Version \textit{6.0.1} was introduced one year after version \textit{5.0.2}. But, version \textit{6.0.1} continues receiving updates 9 months after version \textit{5.0.2}.
    
    %\item All other versions that we take into account, versions 7.0, 7.1.1, 7.1.2, 8.0, 8.1, and 9, are still receiving updates to the time that we have collected our dataset (January 2019).
    
    %\item We see the first update for version 5.1.1 on November of 2015 (6.0 and below) which was released on April 2015 which is a long time. Furthermore, it is interesting to see that the first update for version 6.0 is on November 2015 which is also released at November 2015.
%\end{enumerate}
    
%WHY THIS CLAIM?? Based on the above observations, we observe that Google provides better security update policies in terms of longevity from version to version (from version 4.4.4 to version 5.0.2), but mostly for newer versions. 

For our sample, we can deduce that newer pairs of versions tend to have shorter gaps between EOL times in comparison to their release time gaps. In contrast, service for some older OS versions was abandoned comparatively early (i.e., for 4.4, 5.0, and 5.1), when compared to the gaps in release times with their successors.

Please note that our observations are based on a reduced sample size. On the one hand, AOSP versions 7.0 and higher are still receiving security updates and we cannot make any claim about them. On the other hand, Google started Android security bulletins in August 2015 and does not provide information before that. %As a result, our analysis covers AOSP version 4.4 onwards. 

%As a result, our analysis is limited to from AOSP version 4.4 and afterwards.

%Based on the above observations, we observe that Google provides better security update policies in terms of longevity for newer versions and from version to version (from version \textit{4.4.4} to version \textit{5.0.2}). It is necessary to mention that this observation is based on limited information. In one side, AOSP versions 7.0 and higher are still receiving security updates and we cannot make any claim about them. On the other side, Google have started Android security bulletins on August 2015 and does not provide information before that. As a result, our information are from AOSP version 4.4 and afterwards. 

%Overall, it seems that over time we are seeing better policy. But, it is very questionable that why they have done that. Furthermore, we have only partially have data from version 4.4 and after that. We would have probably seen the same behavior for earlier versions of Android that we do not have data on Android Security Bulletin. 

%In our analysis, we have removed 6.1 from our analysis. We only see two CVEs for that: CVE-2016-2496 and CVE-2016-3843. 

%Android version 4.4 was released on October 31, 2013. In our collected dataset, we do not see any updates for these subversions of 4.4, 4.4.1 (released on December 5, 2013), 4.4.2 (released on December 9, 2013), and 4.4.3 (June 2, 2014). The last subversion of 4.4 was released on June 19, 2014 as 4.4.4. Let's look at each of these versions separately.

\subsubsection{Updated AOSP Versions vs. Affected Versions}
\label{subsub:InConsistent}

In our security bulletin dataset, there are 782 CVEs that have AOSP versions on both the Android Security bulletin and CVEDetails. Among these 782 CVEs,
%\change{of them}{} 
different OS versions are listed on the bulletin and on CVEDetails for 483 of them.
%\Aron{different compared to what? different OS versions are listd on the bulletin and on CVEDetails?} different OS versions.
In this part, we do not focus on the quantitative analysis of these differences between Android security bulletins and CVEDetails. Rather, we focus on two examples and the implications of these differences.

The first example that we study
%\change{take into account}{study} 
is \texttt{CVE-2016-5348}, which
%\change{. This CVE}{, which} 
was mentioned in the October 2016 Android security bulletin \cite{Android_Security_Bulletin_2016_10}. According to the Android security bulletin, the updated AOSP versions are as follows: 4.4.4, 5.0.2, 5.1.1, 6.0, 6.0.1, 7.0. However, according to CVEDetails, the 
%\change{In CVEDetails, there exists a part representing the \textit{affected products}. The}{However, according to CVEDetails, the} 
affected products for this CVE are as follows: 4.0, 4.0.1, 4.0.2, 4.0.3, 4.0.4, 4.1, 4.1.2, 4.2, 4.2.1, 4.2.2, 4.3, 4.3.1, 4.4, 4.4.1, 4.4.2, 4.4.3, 4.4.4, 5.0, 5.0.1, 5.1, 6.0, 6.0.1, 7.0. As we can see, the affected products in CVEDetails are more comprehensive and contain some older versions that are not in the list of updated AOSP versions in Android security bulletins, such as 4.0. Hence, this shows that Google does not provide security updates for all affected versions.\footnote{We assume that information on the Android security bulletins correctly reflects the released patches. It is possible that some information about patched vulnerabilities is not communicated correctly on the Android security bulletins. In particular, that information about released patches for versions that have reached EOL is incorrect. However, we have found no evidence of such practices so far.} The underlying reason is that a version has reached its end-of-life and does not receive security patch updates anymore. Therefore, the older versions of Android are at risk.

The second example is \texttt{CVE-2017-0807}, published in December 2017~\cite{Android_Security_Bulletin_2017_12}. On Android security bulletin, the updated AOSP versions are listed as follows: 5.1.1, 6.0, 6.0.1, 7.0, 7.1.1, 7.1.2. However, in CVEDetails, the affected products are mentioned as follows: 4.0, 4.0.1, 4.0.2, 4.0.3, 4.0.4, 4.1, 4.1.2, 4.2, 4.2.1, 4.2.2, 4.3, 4.3.1, 4.4, 4.4.1, 4.4.2, 4.4.3, 4.4.4, 5.0, 5.0.1, 5.0.2, 5.1, 5.1.1, 6.0, 6.0.1, 7.0, 7.1.1, 7.1.2.
Similar to the first example, Google does not provide security updates for all affected versions that are mentioned on CVEDetails, which
%\change{. This}{, which} 
puts many older devices at risk. 
%Furthermore, these two examples show that \Aron{two examples do not demonstrate that \textbf{many} CVEs are common, they demonstrate only the existence of two CVEs that are}many CVEs are common among many versions, and Google only provides those versions that receive updates rather than the affected versions on the Android security bulletin.

In the following subsection, we investigate the common CVEs among different versions and the possibility that some of them are applicable to older versions.

\subsubsection{Common Vulnerabilities Among AOSP Versions}
\label{subsub:common}

Before starting the analysis of common vulnerabilities among different versions, we introduce
%\change{propose}{introduce} 
the term \textbf{\textit{latent CVEs}} for an AOSP version. The latent CVEs for an AOSP version are
%\change{is the list of}{are} 
all CVEs that affect that version but do not receive security updates like newer versions since the %\change{that}{the} 
AOSP version has reached its end-of-life. 

\begin{figure}[h]
\centering
\includegraphics[width=\columnwidth]{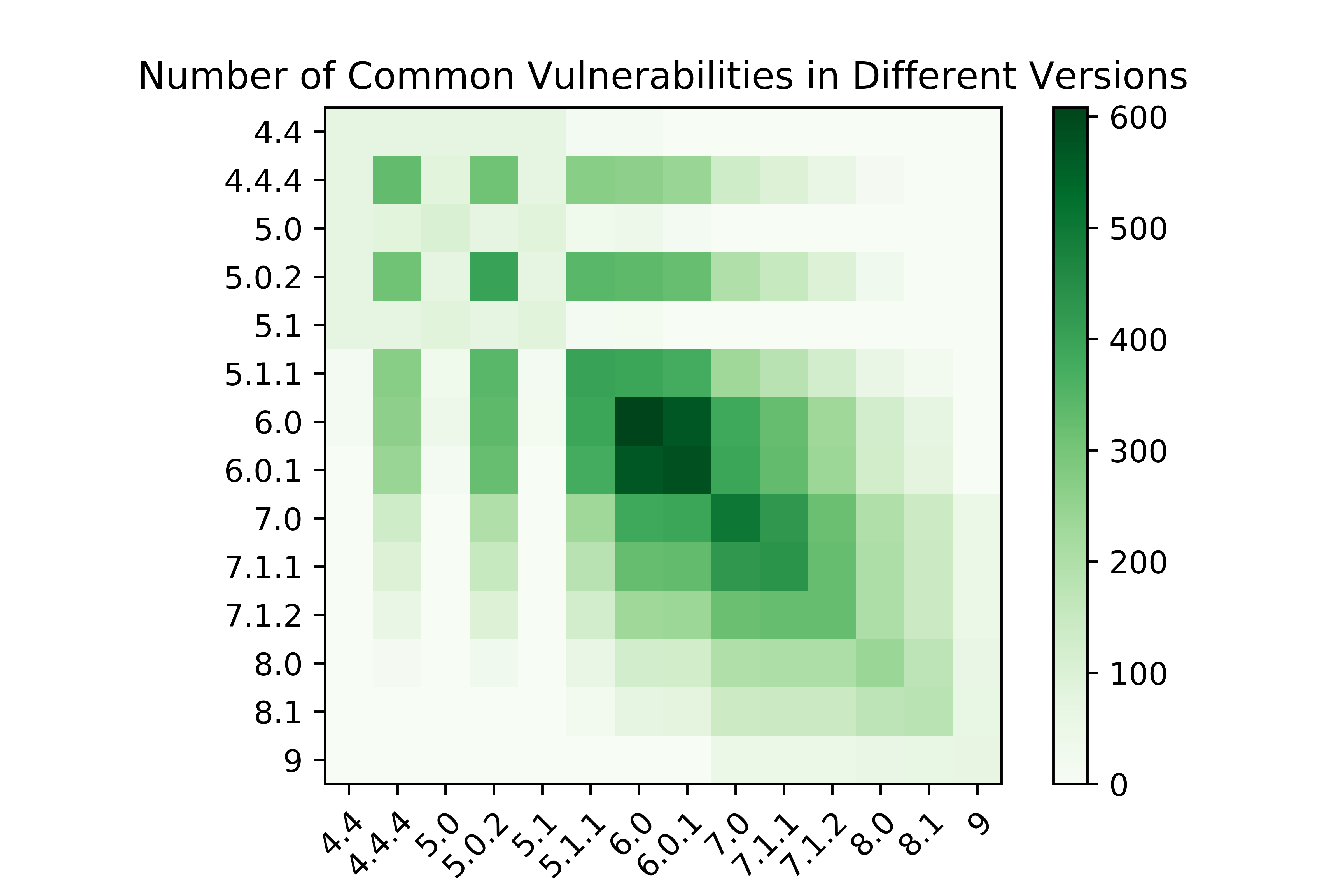}
\caption{Heatmap of common vulnerabilities among the 14 AOSP versions}
\label{fig:all_patched_vulnerabilities_in_2015}
\end{figure}

We analyze the number of common vulnerabilities among different AOSP versions based on the information in Android security bulletins (see Figure~\ref{fig:all_patched_vulnerabilities_in_2015}), however, paying attention to the fact that different versions receive security updates for a different amount of time. The Android security bulletins provide related information about patches, but note that the number of vulnerabilities for different versions varies a lot, see Figure~\ref{fig:number_of_vulnerabilities_affect_each_aops_versions}. As a result, it is highly likely that some Android versions that already receive common patches have even more common vulnerabilities than is reflected in the bulletins. This suggests that many CVEs for a newer version are also applicable for older versions. In the following, we consider the information related to version 4.4 for further analysis.

To understand latent vulnerabilities in version 4.4, we limit ourselves to versions 4.4.4, 5.0, 5.0.2, 5.1, 5.1.1, and 6.0 for comparison purposes. The reason why we do not consider version 6.0.1 and later versions is that we only see updates for version 4.4 until December 2015. Version 6.0.1, on the other hand, was introduced on December 7, 2015. Hence, it is obvious that there should be no common updates between 6.0.1 and versions after that with version 4.4 in our dataset. Note that version 4.4 has 73 patched vulnerabilities. 

Version 4.4.4 has 328 patched vulnerabilities in total and 70 of them are in 2015. All these 70 vulnerabilities are common with version 4.4, while version 4.4 has 3 vulnerabilities which are not reported as common with version 4.4.4. On the other hand, version 4.4.4 continues receiving updates 22 months longer than version 4.4. Therefore, based on our previous discussion and definition of latent CVEs, it is likely that most of the remaining vulnerabilities in version 4.4.4 are also applicable to version 4.4.
    
The above observation might suggest that it is only applicable to sub-versions. To extend our analysis to more significant version changes, we take AOSP 4.4 and 5.x versions as examples. All patched vulnerabilities of version 4.4 are common with version 5.1. With respect to versions 5.0 and 5.0.2, all CVEs are also common (except 1 for 5.0, and 3 for 5.0.2). On the other hand, these three 5.x sub-versions have more vulnerabilities in the observed time period. Therefore, these three sub-versions of version 5 Android have almost all version 4.4 and 4.4.4 vulnerabilities, but also many other CVEs which are not reported as common with the older Android versions (which had an earlier EOL). In other words, version 4.4 CVEs are a subset of the vulnerabilities of versions 5.0, 5.0.2, and 5.1 (while all these versions received updates). This also suggests that many vulnerabilities that were patched later are likely common among versions~4.4 and~5.x.

Based on the exploratory analysis above, many CVEs are likely common among different versions. Once Google stops providing patches for older versions, publishing Android security bulletins provides information to attackers about possible vulnerabilities in these older versions. If the market share of older versions is still non-trivial, this may place many users at risk. %Therefore, it is reasonable to stop security updates for a version when its market share is very low. 

\subsubsection{Unknown CVE ID}
\label{subsub:unknow}

There are 107 CVEs in the Android security bulletins that do not have corresponding entries in CVEDetails. For these CVEs, the CVEDetails website reports ``Unknown CVE ID,'' which means that the corresponding CVE has been reserved, but no information has been provided for it. Other public data sources also return similar results.  For instance, \url{cve.mitre.org} describes these unknown CVE IDs as follows: ``This candidate has been reserved by an organization or individual that will use it when announcing a new security problem. When the candidate has been publicized, the details for this candidate will be provided.'' We believe that this discrepancy is due to the delay in updating these publicly available datasets. To evaluate the role of such delay, we study CVEs in Android bulletins that have unknown CVE IDs. Table~\ref{tab:unknonwn CVE} shows the number of CVEs with unknown CVE IDs for each year from 2015 to 2019. Most of these CVEs are in bulletins from 2018 and 2019. Note that for 2019, we have only the January bulletin. In this month, there are 27 CVEs in total and only 7 of them have corresponding information on the CVEDetails website. Moreover, all of these 7 CVEs are related to the kernel component, which means that these CVEs are not specific to Android but to Linux.
%In other words, the vulnerabilities that originated from Linux community 

\begin{table}[h]
    \centering
    \begin{tabular}{|c|c|c|c|c|c|} \hline
        Year & 2015 & 2016 & 2017 & 2018 & 2019 \\ \hline
        No. Of CVEs & 0 & 5 & 6 & 76 & 20 \\ \hline
    \end{tabular}
    \caption{Number of CVEs with unknown CVE ID in CVEDetails website for each year}
    \label{tab:unknonwn CVE}
\end{table}

%\begin{table}[h]
%    \centering
%%    \begin{tabular}{|c|c|} \hline
%         Layers & No. of CVEs \\ \hline
%         Android runtime & 6 \\ \hline
%         Application & 4 \\ \hline
%         Application framework & 12 \\ \hline
%         Externals & 50 \\ \hline
%         Hardware abstraction & 21 \\ \hline
%         Kernel & 2 \\ \hline
%         Native libraries & 12 \\ \hline
%    \end{tabular}
%    \caption{Number of CVEs and their corresponding layers with unknown CVE IDs}
%    \label{tab:unknown_layer}
%\end{table}

\subsection{RQ3: Vulnerabilities Originating from Qualcomm and Linux}
%How long does it take for Google to handle vulnerability patches of other resources?}
\label{sub:LinQual}
%How long does an external vulnerability patch last for publishing on an Android security bulletin?
%How long does it take for AOSP to handle vulnerabilities of other resources?

%How long does importing external vulnerability patches into AOSP take?}

\begin{figure}[h]
\centering
\includegraphics[width=\columnwidth]{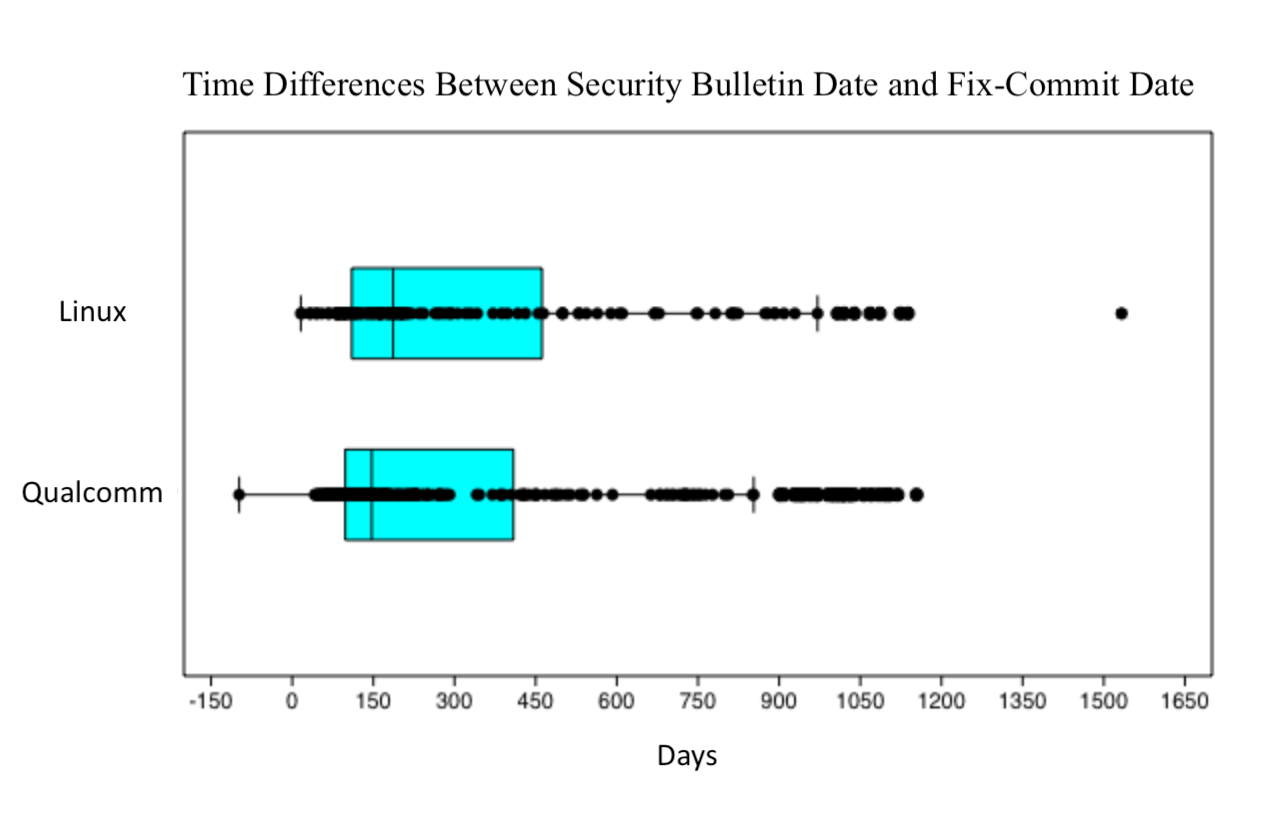}
\caption{Time differences for patched vulnerabilities between the published dates in the Android security bulletins and fix-commit dates of both Qualcomm and Linux patches.}
\label{fig:linux_qualcomm_android}
\end{figure}

To study how long it takes for Google to patch vulnerabilities in components from other vendors and organizations, we focus on kernel and Qualcomm-related vulnerability patches that affect Android. 
In total, there are 1,092 Qualcomm-related and 210 kernel layer patches in the Android security bulletins. However, only 414 Qualcomm and 144 kernel layer patched vulnerabilities have references. For these vulnerabilities, we calculate the time difference of the published date on the respective Android security bulletin and the fix-commit date, which is shown in Figure~\ref{fig:linux_qualcomm_android}.
%because of the security issues, 
%we calculate the time difference of 414 of Qualcomm and 144 kernel layer vulnerability patches. 
%Figure~\ref{fig:linux_qualcomm_android} depicts the time differences of published date on a bulletin and fixing-commit date. 
According to this figure, there is only one patched vulnerability, \texttt{CVE-2017-8281}, whose publication date (September 2017~\cite{Android_Security_Bulletin_2017_09}) precedes its  fix-commit date (December 2017~\cite{Code_Aurora_Minus_Commit}). For all other vulnerability patches, the time difference is positive (see Figure~\ref{fig:linux_qualcomm_android}). In other words, both Qualcomm and Linux patch vulnerabilities before Google publishes them on its security bulletins. The time differences vary mostly between roughly 120 and 450 days (i.e., between 4 months and 15 months). Table~\ref{tab:Mean_Stdev_Linux_Qualcomm} shows the mean and standard deviation of these time differences for both Qualcomm and kernel layer patched vulnerabilities.
The mean values are 307 and 324 days for Qualcomm and kernel layer patches, respectively, and the standard deviations are relatively high for both.
These results indicate that there exists a considerable delay from Google to provide patches for vulnerabilities originated from Qualcomm and Linux. This issue is even more daunting considering that these fix-commit dates are publicly available. Hence, this delay from Google puts many devices at risk. In other words, some of these vulnerabilities originating from Qualcomm and Linux may be interpreted as zero-day vulnerabilities for Android devices.

%Therefore, this tells us that there is a considerably high amount of time for importing the external vulnerability patches into AOSP. Since this can be up to more that one year, a device may be vulnerable during this time. 
%\color[red]{SHOULD WE SAY MORE?}

\begin{table}[h]
    \centering
    \begin{tabular}{|c|c|c|} \hline
        \textbf{Vendor} & \textbf{Mean} & \textbf{Standard Deviation} \\ \hline
        Qualcomm & 307.80 & 323.53 \\ \hline
        Linux & 324.86 & 310,54 \\ \hline 
    \end{tabular}
    \caption{Mean and standard deviation of the time difference between the Android security bulletin time and the fix-commit date for Qualcomm and Linux patched vulnerabilities}
    \label{tab:Mean_Stdev_Linux_Qualcomm}
\end{table}

%The other important issue is that this time distribution differs from Linux to Qualcomm. Analysing this matter is crucial as there maybe different key characteristics that vary from company to company. 

We perform a \textit{Mann-Whitney U} test to see whether there exists a statistical difference between Qualcomm and Linux when Google handles the vulnerabilities originating from them. Since the $p$-value of the test result is 0.033 and is lower than our confidence level which is $p=0.05$, these two distributions are statistically different from each other. Therefore, this shows that Google handles vulnerabilities originated from open source platforms differently from the closed source platforms, i.e., Qualcomm. In this case, the mean time for Linux is higher than for Qualcomm. The underlying reason might be the more open, but diverse, software community in the Linux ecosystem, which corresponds to a longer time for Google to identify and address relevant vulnerabilities. As a result, it is crucial for the Android security team to closely monitor the Linux-related CVEs. 

\subsection{RQ4: Public Disclosure vs Patch Release}
%How long does a vulnerability patch last from its public disclosure to bulletin time?}
\label{sub:delay}

%\textcolor{red}{THINK THE QUESTION ONE MORE TIME}

To investigate this question, we compare three different times. The first one is the \textbf{patch release date}, which indicates when a vulnerability patch is published on an Android security bulletin. The second one is the \textbf{public repository disclosure date} representing when a vulnerability detail is available in public repositories like CVEDetails. The third one is the \textbf{last commit date} indicating when the last fix-commit is made. Having the first commit and the bug creation date would also help us to understand the entire timeline of a vulnerability. However, since these dates are not publicly available, we focus on the aforementioned three dates. 
The relation between these three dates in an ideal setting should be as follows. The first date should be the last commit date, which should then be followed by the patch release date. The patch release date is the point in time where Android releases its security patches. If a public repository disclosure date is sooner than the patch release date, this may place many Android devices at risk. As a result, the public repository disclosure date should be after the patch release date.
%In other words, publicly disclosure date should be always the last date as the public disclosure means that vulnerability details are publicly available from that date. 
%Ideally, vulnerability should be patched and published before the vulnerability details are publicly available.
%Figure~\ref{fig:timeline_commit_bulletin_disclosure} represents the relation among these three time sequences. 

%\begin{figure}[h]
%\centering
%\includegraphics[width=0.8\columnwidth]{images/timeline_commit_bulletin_disclosure.png}
%\caption{The ideal timeline of fix and disclosure of a patched vulnerability}
%\label{fig:timeline_commit_bulletin_disclosure}
%\end{figure}

We limit our analysis to those patched vulnerabilities that have (only) AOSP Git repository references. In other words, we do not consider other external references/repositories like Qualcomm and Linux, since those vulnerabilities are patched in other repositories in addition to AOSP Git repositories.
%Since, those vulnerabilities are published in other resources in addition to Android security bulletin.
Note that a patch release date does not exactly correspond to the time of disclosing a patched vulnerability. Only 632 patched vulnerabilities fall into this criterion, i.e., having AOSP Git repository references. Also note that a patched vulnerability can have multiple references. Therefore, the time sequence can change since each reference has its own last fix-commit date. 

Given that, we separately analyze the following two different groups: Patched vulnerabilities that have only one reference and the others that have more than one reference. For the public repository disclosure date, we use the date that is indicated as \textit{publish date} on the CVEDetails web page. Our level of granularity to capture these three dates are year and month. Therefore, we do not consider the exact calendar day of release/publication in our analysis. 
%Table~\ref{tab:frequency_only_one_aosp} shows the frequency of the all time orders seen so far. 

In our analysis, we use the symbols  $\mathbb{B}$, $\mathbb{C}$ and $\mathbb{D}$ to denote the \emph{patch release date}, \emph{fix-commit date}, and \emph{public repository disclosure date}, respectively.
%In our analysis, we use the following notations. $\mathbb{B}$ denotes the bulletin date. The fix-commit date is represented by $\mathbb{C}$ we denote public repository disclosure date by $\mathbb{D}$.
To express the sequence of these three dates, we separate them with ``-'' if they happened at different times. The date on the left side of ``-'' is earlier than the date on the right side.
%Note that the timing order is from left to right. 
For example, $\mathbb{C-B-D}$ means that the commit occurred first; then, the patch was released, which was later followed by public repository disclosure. If two dates are not separated by ``-'' (e.g., $\mathbb{BC}$), then they are in the same month. For instance, $\mathbb{BC-D}$ means that patch release and last commit are in the same month, and the public repository disclosure is after them. 
%Note that we only count years and months, not days because scraping the exact bulletin date with having the day needs further work. Hence, the cases like BC-D, we only know that both bulletin and commit date appears in the same month. 
Table~\ref{tab:frequency_only_one_aosp} shows the frequency of all the time sequences that we observe in the Android ecosystem.

\begin{table}[h]
    \centering
    \begin{tabular}{|c|c|c|} \hline
        \textbf{Time Sequence} & \textbf{Frequency} \\ \hline
        $\mathbb{BC-D}$ & 1 \\ \hline 
        $\mathbb{C-B-D}$ & 66 \\ \hline 
        $\mathbb{C-D-B}$ & 3 \\ \hline 
        $\mathbb{C-BD}$ & 530 \\ \hline 
        $\mathbb{CBD}$ & 2 \\ \hline 
        $\mathbb{D-C-B}$ & 19 \\ \hline 
        $\mathbb{DC-B}$ & 11 \\ \hline 
    \end{tabular}
    \caption{The frequencies of the time sequences of  patched vulnerabilities that have only one AOSP reference}
    \label{tab:frequency_only_one_aosp}
\end{table}

According to Table~\ref{tab:frequency_only_one_aosp}, of 632 patched vulnerabilities, only 66 followed the ideal time sequence $\mathbb{C-B-D}$. However, 530 of them follow a near ideal sequence, the only difference being that the patches are released \textit{and} publicly disclosed in the same month. Therefore, 94\% of the vulnerabilities are patched no later than their disclosure date in public repositories. %before or the same date with publicly disclosure date. 
The remainder, on the other hand, have different timelines. The most problematic one is the vulnerability being disclosed in public repositories and then receiving a patch, which is represented by $\mathbb{D-C-B}$ and $\mathbb{DC-B}$ in our dataset (4\%). For instance, \texttt{CVE-2017-6983} was published in September 2017~\cite{Android_Security_Bulletin_2017_09} and its last commit date is on August 2018. However, it was disclosed in public repositories on May 2017. Similarly, \texttt{CVE-2017-13078} was published in November 2017~\cite{Android_Security_Bulletin_2017_11}, but its disclosure date in public repositories and its last commit data are in % the same time, i.e., 
October 2017.

Similarly, Table~\ref{tab:frequency_more_than_one_aosp} shows the frequencies of time sequences for patched vulnerabilities that have more than one AOSP reference. In other words, in this table, each patched vulnerability has more than one last commit date. Based on Table~\ref{tab:frequency_more_than_one_aosp}, the majority of the patched vulnerabilities still follow the same timeline. Sequences $\mathbb{C-BD}$ and $\mathbb{C-B-D}$ contain 304 out of 321, i.e., 94\% of total commits that belong to 125 different patched vulnerabilities. This shows that for the majority of patched vulnerabilities we do not see potentially dangerous practices. Particularly, the disclosure date is not sooner than the patch time. 
%There are still, on the other hand, exceptions though. The number of $\mathbb{D-C-B}$ and $\mathbb{DC-B}$ is 16. Similarly, 
\texttt{CVE-2014-6060}, published in April 2016, is the one vulnerability corresponding to $\mathbb{C-D-B}$~\cite{Android_Security_Bulletin_2016_04}.
%which is CVE-2014-6060. The bulletin date of this vulnerability is April 2016~\cite{2016-4}\textcolor{red}{add citation}, 
This vulnerability was disclosed in public repositories in September 2014, and one of its two references has June 2014 as the last commit date. The other reference is in September 2014, which is the same as the public repository disclosure date. In addition, vulnerability \texttt{CVE-2014-6060} is the only one that fits more than one time sequence. Since the first commit is in June 2014, it fits $\mathbb{C-D-B}$. It also fits $\mathbb{CD-B}$ because the second commit is in September 2014, which is the same as its public repository disclosure date. The other vulnerabilities fit only one time sequence each. 

\begin{table}[h]
    \centering
    \begin{tabular}{|c|c|c|} \hline
        \textbf{Time Sequence} & \textbf{Frequency} \\ \hline
        $\mathbb{C-B-D}$ & 34 \\ \hline
        $\mathbb{C-D-B}$ & 1 \\ \hline
        $\mathbb{C-BD}$ & 270 \\ \hline
        $\mathbb{D-C-B}$ & 15 \\ \hline
        $\mathbb{DC-B}$ & 1 \\ \hline
    \end{tabular}
    \caption{The frequencies of the time sequences among the patched vulnerabilities that have more than one AOSP reference}
    \label{tab:frequency_more_than_one_aosp}
\end{table}

\begin{figure}[h]
\centering
\includegraphics[width=\columnwidth]{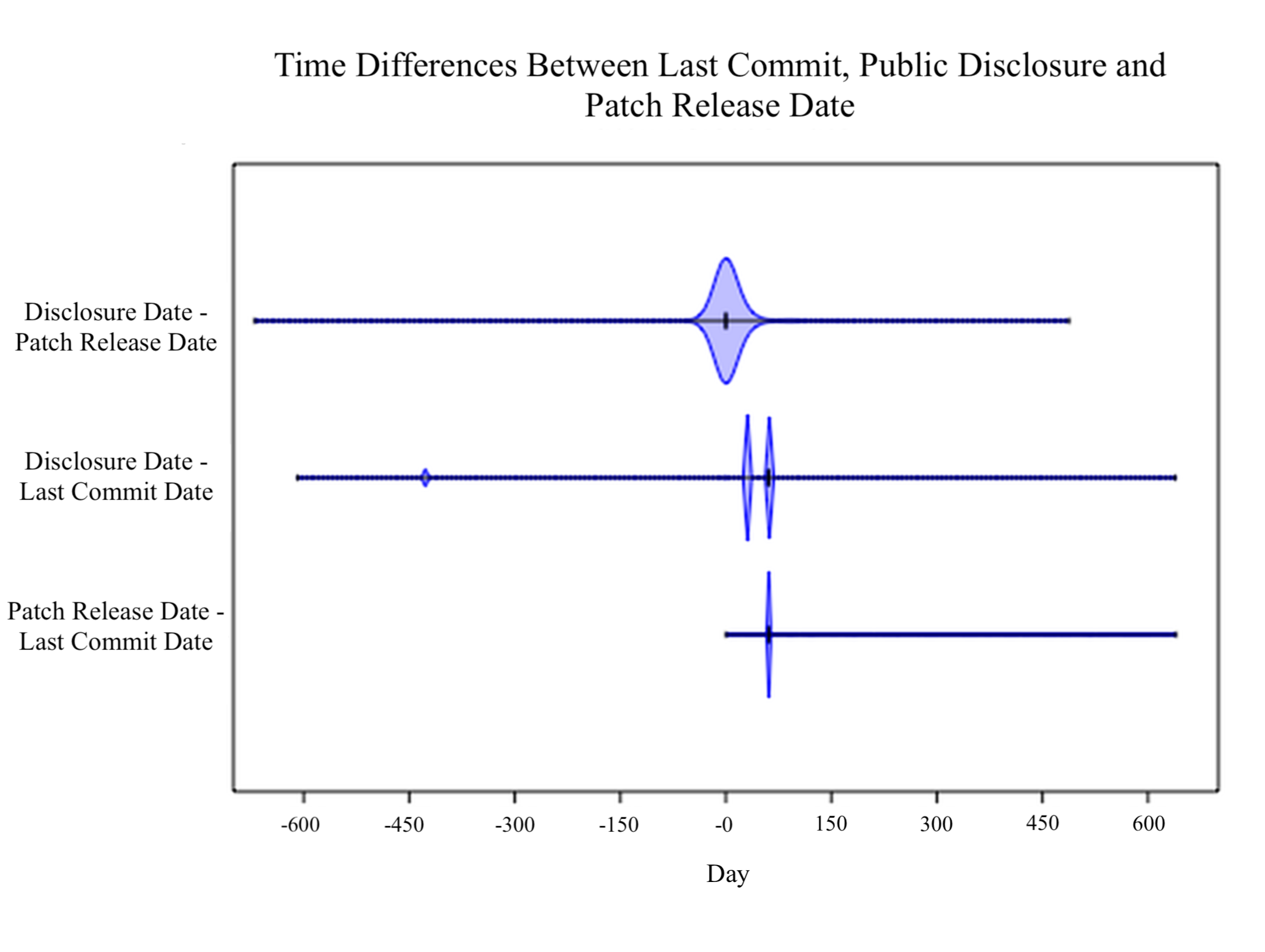}
\caption{Time gaps between the last commit, publicly disclosure and patch release dates}
\label{fig:disclosure_commit_bulletin}
\end{figure}

Considering these two tables, the majority falls into $\mathbb{C-B-D}$ and $\mathbb{C-BD}$ which is a secure practice. However, there are some exceptions. For example,  %Figure~\ref{fig:timeline_commit_bulletin_disclosure}. 
\texttt{CVE-2014-6060} follows the following two time sequences: $\mathbb{C-D-B}$ and $\mathbb{CD-B}$. For this patched vulnerability, the difference between disclosure date in public repositories and patch release date is around 1.5 years. However, these two aforementioned tables do not indicate the actual time gaps. Therefore, we further analyze the last commit, public repository disclosure, and patch release dates. The main goal is to determine the distribution of time gaps between these three points in time. Figure~\ref{fig:disclosure_commit_bulletin} shows these dates for all patched vulnerabilities that have public repository disclosure dates and have at least one AOSP reference. Note that, if there is more than one reference, we calculate the time differences for each of them. For example, if a patched vulnerability has two references, there are two different time results that specify the comparison of both patch release date and disclosure date in the public repository. Table~\ref{tab:disclosure_commit_bulletin_mean_stdev} shows the means and standard deviations of the results represented in Figure~\ref{fig:disclosure_commit_bulletin}. For the analysis of \textbf{disclosure date of the public repository and patch release date}, we analyze 758 patched vulnerabilities. For both analysis of \textbf{patch release date - last commit} date and \textbf{disclosure date of public repository - last commit date}, we analyze 954 patched vulnerability references. The reason for having more patched vulnerabilities for the last two analyses is due to multiple references for some patched vulnerabilities. 

\begin{table}[h]
    \centering
    \footnotesize
    \begin{tabular}{|c|c|c|} \hline
        \textbf{Time Gap Analyze} & \textbf{Mean} & \textbf{Standard Deviation} \\ \hline
        Disclosure Date - Patch Release date & -2.43 & 68.2071 \\ \hline
        Disclosure Date - Last Commit Date & 54.26 & 81.5345 \\ \hline
        Patch Release date - Last Commit Date & 61.84 & 43.7562 \\ \hline
    \end{tabular}
    \caption{Means and standard deviations of time differences among patch release date, last commit date, and disclosure date in public repository}
    \label{tab:disclosure_commit_bulletin_mean_stdev}
\end{table}

According to the Figure~\ref{fig:disclosure_commit_bulletin}, there are no negative values for the time differences between the patch release date and the commit date. In other words, patch release dates occur after the commit date for all patched vulnerabilities. This is  expected since a patch release date indicates when a patch is available for a particular vulnerability; it should also occur after its last fix-commit date. The mean of this time gap is 61.84 days. In other words, for publishing a vulnerability patch after making the last fix-commit, Google spends an average of around 62 days (2 months). However, there are some extreme cases. The longest time gap is 640 days which belongs to \texttt{CVE-2014-6060}, which we already mentioned above. The second longest time gap is 639 days belonging to \texttt{CVE-2016-1621}. It is published in March 2016~\cite{Android_Security_Bulletin_2016_03} and disclosed in a public repository in the same month. However, it has 3 different references. Although two of them have the same last commit dates (January 2016), the third one has a date of June 2014.

For the time differences between the disclosure date in a public repository and the patch release date, even though we see large variation, from lower than -600 days to around 488 days, the majority of them, i.e., for 639 vulnerabilities, occurs at the same time. In other words, 84\% of them are disclosed at the same time in a security bulletin as in a public repository. 10\% of them have positive values which means that they are disclosed in public repositories after the patch release date. In total, around 95\% of them are disclosed in a public repository after or at the same date as the patch release date. The mean of this time difference is equal to -2.43, i.e., very close to zero. Around 5\% of them have a negative value. In other words, they are publicly disclosed before Android published them on its security bulletins, which is the same date when Google provides its patches. Even though this occurs for a limited number of vulnerabilities, it is a potentially dangerous case which puts many devices at risk considering that a vulnerability has been published publicly but the patch has not been provided.

Lastly, we investigate the time difference of disclosure time in a public repository and the last commit date. 94\% of them have a positive value meaning that they are publicly disclosed in a repository after the last commit is made. For 66\% and 12\% of our total samples, the time differences are 60 and 30 days, respectively, which we can also see in Figure~\ref{fig:disclosure_commit_bulletin}. The mean value for this time difference is equal to 54.26. Hence, it takes around 2 months for public repositories to disclose a vulnerability after its last commit date.
%we can see that almost all of the vulnerabilities are publicly disclosed after around 2 months where the last commit is made. 
Note that for 3\%, i.e., 34 of them, the disclosure time in a public repository is even sooner than the last commit date. This means that a vulnerability has been disclosed in a public repository while the fix has not been yet finalized. This gives an attacker an advantage to compromise Android devices. 

Based on our above analysis, for the majority of vulnerabilities, these three time sequences follow the pattern which can be considered secure. However, there are some instances that put the security of Android devices at risk. All of these instances show that a vulnerability has been introduced in public repositories when the patch is not available or even before the last fix-commit in Google Git. Even though these instances are not common, i.e., about 5\%, the potential consequences might be significant. It follows that we need better coordination between Google and public repositories with respect to their time management in announcing vulnerabilities. %For a secure Android ecosystem, it is crucial a public repository does not publicize a vulnerability before its patch is made available by a vendor. 

%As a result, the majority of the vulnerabilities are publicly disclosed at least the same date with their bulletin dates.
%\textcolor{red}{SAY MORE?}

\subsection{RQ5: Vulnerability Lifetime}
%How long does a vulnerability take to patch?}
%What are the survivability days of patched vulnerabilities?
%}

Vulnerability lifetime means the time difference between the introduction of vulnerability in the code and when the vulnerability is eventually patched~\cite{linares2017empirical}. 
It is difficult to find an actual patching time for Android vulnerabilities, i.e., when Google releases the security patches for a vulnerability. For this reason, the published time of a security bulletin gives the upper bound for when a patch is released; and we consider this as the patching date. In order to calculate the introduction time of a vulnerability in the code, we use an algorithm called SZZ~\cite{SZZ}. The general idea behind the SZZ algorithm is to identify the changed lines for fixing a bug and then identify when these lines were added for the first time. By doing so, first, we manually clone all the AOSP Git repository branches being used for patching the vulnerabilities.

After cloning the branches, for each reference, we take the commit ID. With this commit ID, we issue the command \textit{diff}~\cite{Git_Diff} under the corresponding branch in order to detect which lines are changed from the previous revision to last revision. Since the algorithm relies on only the line deletions, we have to find when these deleted lines were added in the first place. Hence, we use the command \textit{annotate} on the previous revision for each changed file where the changes were made ~\cite{Git_Annotate}. Following this process, we can find the first time when the lines were added to the code. Note that there can be more than one deleted line. Furthermore, it is possible that each deleted line is added at different times. As a result, we use the terms \textbf{\textit{maximum lifetime}} and \textbf{\textit{minimum lifetime}} for better understanding of this time interval. These terms are in agreement with the original SZZ paper~\cite{linares2017empirical}. The maximum lifetime shows the time interval between the line addition time of the first deleted line and the time of publishing the vulnerability patch on the security bulletin. On the other hand, the minimum lifetime is the time interval between the line addition time of the last deleted line and the time of publishing the vulnerability patch on the security bulletin.

\begin{figure}[h]
\centering
\includegraphics[width=\columnwidth]{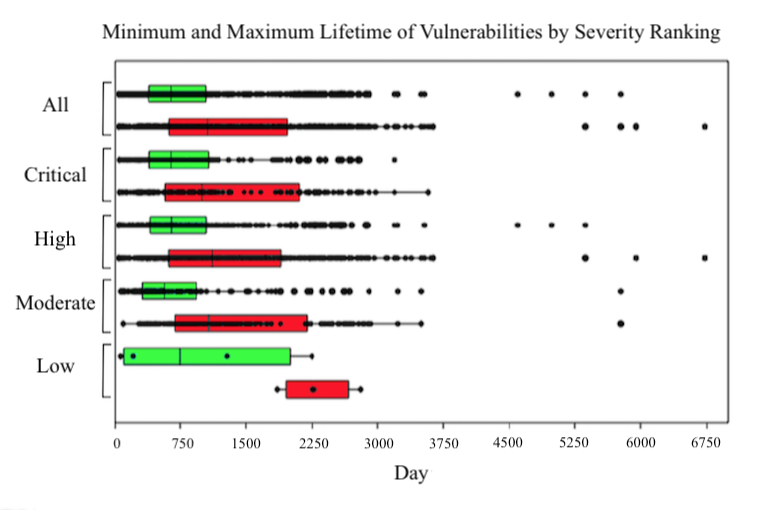}
\caption{Green and red bar represents the minimum and maximum vulnerability lifetime, respectively}
\label{fig:max_min_survivability}
\end{figure}

As mentioned before, this algorithm only checks the deleted lines. As a result, it excludes the cases where only additions are made. To identify that, a static code analysis must be done which is not the focus of this work. Besides, we also exclude the patched vulnerability that do not have any references. After excluding all of them, we have 549 patched vulnerabilities for which we are able to identify the maximum and minimum lifetimes. Among them, we notice 133, 300, 112, and 4 patched vulnerabilities with critical, high, moderate, and low severity levels, respectively. The result of our analysis for both maximum lifetime and minimum lifetime is represented in Figure~\ref{fig:max_min_survivability}. %Further, the result for all of the patched vulnerabilities are plotted on top of all severity levels. 

As we see in Figure~\ref{fig:max_min_survivability}, both minimum and maximum patched vulnerability lifetimes are very high. In particular, there are  patched vulnerabilities with high and low severity levels that have a maximum vulnerability lifetime of more than 6000 days, which is longer than the lifetime of Android. The reason is that the algorithm also checks files and lines that might be irrelevant for the vulnerability like a log, build files and even comments. For instance, if the fix-commit includes a comment deletion, then the algorithm takes this into consideration. Therefore, the above outcomes can occur, for example, when a build file or log file has not been changed for 7-8 years and then has been changed in the last fix-commit. Further, since this issue might also happen on external branches that have older commit histories than Android AOSP, we can observe these outliers. In general, the minimum patched vulnerability lifetimes fluctuate between 300 and 800 days, which is still high. On the contrary, the maximum vulnerability lifetimes vary from 700 to 2200 days (including the aforementioned outliers).

Table~\ref{tab:mean_stdev_max_vulnerability} and Table~\ref{tab:mean_stdev_min_vulnerability} report the mean, and standard deviation results for both maximum and minimum vulnerability lifetimes.

\begin{table}[h]
    \centering
    \begin{tabular}{|c|c|c|} \hline
        \textbf{Severity Rankings} & \textbf{Mean} & \textbf{Standard Deviation}  \\ \hline
        All Vulnerabilities & 1350.2 & 981.69 \\ \hline
        Critical & 1254.49 & 860.21 \\ \hline
        High & 1366.49 & 1042.53  \\ \hline
        Moderate & 1386.66 & 949.10 \\\hline
        Low & 2295.25 & 391.20 \\ \hline
    \end{tabular}
    \caption{Mean and standard deviation values of maximum vulnerability lifetimes}
    \label{tab:mean_stdev_max_vulnerability}
\end{table}

\begin{table}[h]
    \centering
    \begin{tabular}{|c|c|c|} \hline
        \textbf{Severity Rankings} & \textbf{Mean} & \textbf{Standard Deviation} \\ \hline
        All Vulnerabilities & 882.78 & 789.59 \\ \hline
        Critical & 868.53 & 697.47 \\ \hline
        High & 890.68 & 796.92  \\ \hline
        Moderate & 876 & 871.68 \\ \hline
        Low & 954 & 1018.65 \\ \hline
    \end{tabular}
    \caption{Mean and standard deviation values of minimum vulnerability lifetimes}
    \label{tab:mean_stdev_min_vulnerability}
\end{table}

%For high and low severity level patched vulnerabilities, we realize maximum vulnerability lifetime which is more than 6000 days, which is more than the lifetime of Android.

In order to see whether these data are statistically different, we perform a \textit{Mann Whitney U} test for all minimum and maximum vulnerability lifetime data with each other. In other words, each minimum/maximum vulnerability lifetime set is compared to another dataset of a minimum/maximum vulnerability lifetime with different severity levels. Our null hypothesis, $H_O$, is that all minimum/maximum vulnerability lifetime datasets are equally distributed. 
%With given the p //SPECIAL CHARACTER value 0.05, 
Table~\ref{tab:Max_Whitney} and Table~\ref{tab:Min_Whitney} depict the $p$-values of the Mann-Whitney U tests.

%In order to see whether these datasets are statistically different, we use \textit{Mann Whitney U} test. We use Mann Whitney U test several times for all minimum vulnerability lifetime datasets as well as maximum vulnerability lifetime. In other words, each minimum (maximum) vulnerability lifetime test is compared with another dataset of minimum vulnerability lifetime with different severity level. Our null hypothesis, $H_O$, is that all minimum vulnerability lifetime datasets are equally distributed. 
%With given the p //SPECIAL CHARACTER value 0.05, Table~\ref{tab:Max_Whitney} and Table~\ref{tab:Min_Whitney} depict the $p$-values of the Mann-Whitney U tests.

According to the results, the $p$-values of all pairs are larger than the significant level, i.e., $p$ = 0.05. This means that the null hypothesis cannot be rejected. In other words, there are no statistical differences between the datasets. 

\begin{table}[h]
    \centering
    \begin{tabular}{|c|c|c|} \hline
        \textbf{Maximum Lifetime} & \textbf{U-value} & \textbf{$p$-value} \\ \hline
        Critical-High & 18994.5 & 0.497 \\ \hline 
        Critical-Moderate & 6760.0 & 0.204 \\ \hline
        Critical-Low & 99.5 & 0.396 \\ \hline
        High-Moderate & 16267.5 & 0.168 \\ \hline
        High-Low & 203.5 & 0.399 \\ \hline
        Moderate-Low & 90.0 & 0.375 \\ \hline
    \end{tabular}
    \caption{Comparison results of Mann Whitney U test performed on maximum lifetime values for each severity levels}
    \label{tab:Max_Whitney}
\end{table}

%\begin{figure}[h]
%\centering
%\includegraphics[width=\columnwidth]{images/Maximum_mann_whitney.png}
%\caption{p //SPECIAL CHARACTER values for comparison of maximum survivability sets}
%\label{fig:maximum_mann-whitney-u-test}
%\end{figure}

\begin{table}[h]
    \centering
    \begin{tabular}{|c|c|c|} \hline
        \textbf{Minimum Lifetime} & \textbf{U-value} & \textbf{$p$-value} \\ \hline
        Critical-High & 19943.0 & 0.497 \\ \hline 
        Critical-Moderate & 6992.0 & 0.204 \\ \hline
        Critical-Low & 245.0 & 0.346 \\ \hline
        High-Moderate & 15765.0 & 0.168 \\ \hline
        High-Low & 555.0 & 0.399 \\ \hline
        Moderate-Low & 202.5 & 0.375 \\ \hline
    \end{tabular}
    \caption{Comparison results of Mann Whitney U test performed on minimum lifetime values for each severity levels}
    \label{tab:Min_Whitney}
\end{table}

%\begin{figure}[h]
%\centering
%\includegraphics[width=\columnwidth]{images/minimum_mann_whitney.png}
%\caption{p //SPECIAL CHARACTER values for comparison of minimum survivability sets}
%\label{fig:minimum_mann-whitney-u-test}
%\end{figure}

%-------------------------------------------------------------------------------

%-------------------------------------------------------------------------------
\section{Discussion}
\label{sec:Discuss}

We believe that our work represents important steps in understanding the  security practices in the Android ecosystem, as well as its likely impact on users. We now present the emerging themes and practical security policy recommendations based on our study. 

\subsection{Comprehensive Security Bulletins}

The first commercialized version of Android was released on September 23, 2008. Since then, Google has released 63 different AOSP versions with 28 API levels~\cite{All_AOSP_Versions}. Due to the openness of the platform, Android has been adopted by different vendors, like Samsung, LG, etc., which results in the highest market share among mobile phone devices. Considering the large and widespread use of Android devices and their impact on humans lives, it is vital to keep Android devices secure over time. Because of that, Google provides monthly security patch updates to fix security vulnerabilities. Based on data from the CVEDetails website~\cite{androidCVEdetails}, the Android-related CVEs existed back in 2009. However, Google started its Android security bulletins only in August 2015. Subsequently, also some vendors began releasing their own security bulletins. For instance, Samsung started in October 2015 and LG started in May 2016. 

According to the CVEDetails website, there are 2,146 Android-related vulnerabilities by the end of 2018. However, the recorded number of Android-related vulnerabilities from 2009 to the end of 2014 is only equal to 43~\cite{androidCVEdetails}. That means, from 2015 (start of Android security bulletins), the number of Android CVEs has increased drastically. Therefore, it seems that releasing Android security bulletins has certainly provided better knowledge about Android security. In contrast, we do not have too much information about the early days of Android (2008) until 2015. From a research perspective, it would be useful if Google starts adding also security bulletins that belong to the time before August 2015 to enable a comprehensive overview of security patches from the introduction of Android to the present day.

%Whole period of Android rather than just from Aug 2015 to now

%Know the exact date that provide patch

%Why postponing release of external one

\subsection{Coherent Security Platforms}

There are many different platforms providing detailed information of CVEs, like CVEDetails~\cite{Cve_Details}, MITRE~\cite{cve_mitre} and the National Vulnerability Database (NVD)~\cite{National_Vulnerability_Database}. These three are general purpose databases and are not limited to one vendor or platform. Therefore, we can potentially find any vendor-related as well as Android-related vulnerabilities on them. In this paper, we use CVEDetails and Android security bulletins to investigate Android patched vulnerabilities. The expectation is that all of these publicly available platforms should be consistent with each other; and that there are no contradictions. In other words, only checking one of them should be sufficient for the majority of problems. 

However, based on our results in Section~\ref{subsub:InConsistent}, we notice inconsistencies between the Android security bulletins and CVEDetails. These inconsistencies included but are not limited to different information regarding updated and affected AOSP versions, and unknown CVE IDs on the CVEDetails website. Due to these inconsistencies, for example, if a person only relies on CVEDetails, s/he might miss some of the Android-related vulnerabilities on Android security bulletins considering that they are unknown in CVEDetails. Furthermore, Google provides updated versions while CVEDetails provide affected versions. 
%Google does not provide affected versions while CVEDetails provides.
In some cases, these two seem to systematically differ from each other. As a result, an observer cannot rely on only one of them. We believe it is essential that different publicly available websites that explain the vulnerabilities in detail should strive to be consistent with each other. %In other words, each of them should should strive to provide the same information. 
Consistency would help security practitioners to find reliable information easily and they would not have to check different resources. %Furthermore, these inconsistencies might lead to uncertainty since a person cannot check which resource has the missing attribute if there exists an inconsistency for it. 

%Furthermore, these inconsistencies may lead to uncertainty since a person cannot check which resource is the right one if there exist an inconsistency for an attribute. .

\subsection{User-Centric Security Policy and Security after End-Of-Life (EOL)}

Each vendor stops providing software updates after some period of time. Microsoft clearly states the end of support for its Windows operating systems~\cite{Windows_lifecycle_fact_sheet}. For example, for Windows 7, the end of mainstream support was January 13, 2015, and the end of extended support is January 14, 2020. Microsoft provides the exact date of support clearly on their website. Unlike Microsoft, Google does not mention these dates exactly ~\cite{update_nexus_device}; however, clearly announcing such a schedule would be a desirable practice for developers that work within the Android ecosystem as well as consumers. %We believe that it might be more clear for users to know the exact time of EOL rather than comparing two times. Therefore, it would be better to have the exact time. %It might also be better, like Microsoft, to have different phases for EOL. 

%Based on the above policy of Google, it does not consider the consumers at all. 
%Based on the above policy of Google, this policy does not into account consumers at all.
Further, there are still a considerable number of consumers who use Android devices that do not receive security updates anymore. By taking our analysis and results in Section~\ref{sub:eol} into account, many CVEs are common among different versions including versions that have reached EOL. As a result, out-of-date versions are vulnerable as they do not receive security patch updates. Hence, we believe that in order to have a more secure Android ecosystem, the EOL policy should take the consumers more directly into account. Our proposition is that the EOL date should be related to the market share of an Android version. For example, if the market share of an Android version (e.g., 4 or 4.x) drops below a certain threshold, then an EOL date can be announced and notifications should be sent to the users. The other advantage of our proposed policy is that when a market share of an Android version is lower than a threshold, it might not be beneficial (in terms of cost-benefit analysis) for rational attackers to exploit a vulnerability.
%Taolk about the issues of end-of-life and its impact on consumers security. There are multiple considerations. First, Google should be more specific. Gives the exact date rather than two options in describing the actual date. The second idea is that it might make sense to change the policy. The current focus is on time after the release. But, why it should be about time? It may be better to change it to market share and respect more consumers. 

%Google stops providing security updates for each version after some time. For instance, 

%\subsection{User-Centric Security Policy}

%-------------------------------------------------------------------------------
\section{Limitations}
\label{sec:Limit}

%While answering the research questions, we also face limitations. 
In the following, we discuss limitations of our work. 
%The first limitation is lacking of having standardized data in Android security bulletins. As an example, some of the values of column \textit{Updated AOSP Versions} have the terms \textit{below} and \textit{above} while for other vulnerability patches, they are explicitly indicated. Besides, we do not find any satisfying explanation of what these terms may refer to.
The first security bulletin was published in August 2015 by Google while Android was commercialized in 2008. As a result, our analysis and investigation are limited to the last four years (version 4.4 and above) and cannot be extended to 2008-2015. However, we believe that our study and analysis are representative of the current practices of Android and can lead to better policy design in this ecosystem. 
%However, since Google does not provide any data of vulnerabilities patched before August 2015, this is a limitation for us. Especially analyzing the AOSP versions, we have to narrow our analysis down since we do not have any information related to previous AOSP versions that was used in before August 2015.

Another limitation of our work is that we have asymmetric information for different CVEs. For example, for some CVEs in Android security bulletins, we have a reference link which gives some information about the commit date, etc. But, for other CVEs, there are no such references. Another example is the \textit{reported date}. Only a limited number of CVEs have this attribute. As a result, our analysis for these attributes works with a reduced sample size of those CVEs that have that information.

The SZZ algorithm has its own limitations as well. First, the algorithm only considers code deletions when it tries to find the first commit that might cause a particular vulnerability. However, there are other commits that only have code additions and analyzing code additions needs static code analysis to find the cause of vulnerabilities. Furthermore, the SZZ algorithm also checks the files that might not cause a vulnerability. Log files and build files are examples of them. Since there might be lots of different types of files to exclude, we did not do anything further in this case and leave a more nuanced analysis to future work.

The algorithm also proposes to draw information from the issue tracker~\cite{SZZ} such as bug creation date. However, while there is an Android issue tracker~\cite{Android_Issue_Tracker}, we were not able to identify relevant information about patched vulnerabilities. Therefore, it is very likely that these bugs are internally tracked.

%Another aspect that the algorithm proposes but we cannot check is the bug creation date. Taking bug creation dates into account significantly helps to identify the commits that are patched before the bug is created in an issue tracker, as the algorithm indicates~\cite{SZZ}.
%The algorithm indicates that considering the bug creation date helps better to identify the commits that are patched before the bug is created in an issue tracker~\cite{SZZ}.
%However, although there is an Android issue tracker~\cite{Android_Issue_Tracker}, we do not find any issues related to patched vulnerabilities. Therefore, it is very likely that these bugs are internally tracked. 

%-------------------------------------------------------------------------------
\section{Conclusion}
\label{sec:conclusion}

Our paper provides, what we believe to be the most detailed and comprehensive  study on patched Android vulnerabilities. We have collected 2,470 vulnerabilities with their detailed information from both Android security bulletins and the CVEDetails website for August 2015 to January 2019. 

First, we investigated the overall trend of the patched vulnerabilities by analyzing the number of vulnerabilities per year for each severity level. Second, we analyzed the distribution of the root causes of patched vulnerabilities by studying their common weakness enumerations (CWE). Third, we studied the duration of support for different AOSP versions. Fourth, we calculated the time gap between the fix-commit and the published date of vulnerabilities that originate from Linux and Qualcomm. Fifth, we examined the sequences of the public disclosure, the patch release, and the last commit dates of the patched vulnerabilities. Last, but not least, we analyzed the  maximum and minimum vulnerability lifetimes.

For example, our findings demonstrate that the most common root cause of patched vulnerabilities is \texttt{CWE 119: Failure to Constrain Operations within the Bounds of a Memory Buffer}. %Further, the severity level of most patched vulnerabilities. 
%vulnerabilities that have the severity level of the high are the most patched ones for many months.
Further, we showed that the length of security support varies for different AOSP versions. In addition, there are versions that are affected by Android-related vulnerabilities but not updated due to Google's patch policy, which leaves users unprotected after an Android version has reached EOL status. 

We hope that our research contributes to a better understanding of the security practices in the Android ecosystem, and helps to develop better policies for security management in the future.

%We indicate that the time gap between the Android security bulletin time and the fix-commit date for Qualcomm and Linux patched-vulnerabilities are 307 and 324 days, respectively. Also, although the most patched vulnerabilities fixed, patched and publicly disclosed, there are other ones that are publicly disclosed and then are started to patch which puts the Android devices at risk. Our final finding is that the patched vulnerabilities have the 882 and 1350 days for minimum and the maximum lifetimes, respectively. 

%-------------------------------------------------------------------------------
\textbf{Acknowledgments:} We thank the anonymous reviewers for their constructive comments and feedback. We further want to thank Ikra Gizem Yildiz and Ece Kubilay for their feedback. This work was supported by the German Institute for Trust and Safety on the Internet (DIVSI).
%We want to thank the reviewers \Aron{How do we know who the reviewers are?} Ikra Gizem Yildiz and Ece Kubilay for their valuable contributions.
%The research activities of Jens Grossklags are supported by the German Institute for Trust and Safety on the Internet (DIVSI).

%-------------------------------------------------------------------------------
\bibliographystyle{plain}
\bibliography{ref}
\appendix

\section{Android Stack Layers}
\label{app:layers}
%CLASSIFICATION OF VULNERABILITIES INTO ANDROID STACK LAYERS}

\begin{figure}[h]
\centering
\includegraphics[width=\columnwidth]{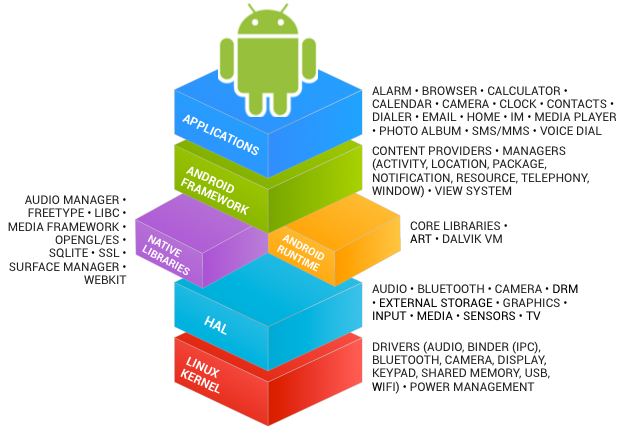}
\caption{Android stack layers~\cite{Android_Stack_Layers}}
\label{fig:android_stack_layers}
\end{figure}

Android stack layers are software components for a wide array of devices with different form factors. It helps understanding where a software component is located. Figure~\ref{fig:android_stack_layers} represents all of its five main stack layers and one additional component~\cite{Android_Stack_Layers}. In this section, we investigate the role of Android stack layers in vulnerabilities. 

It is crucial to examine that which Android stack layer is affected by which vulnerability. Thus, one can see which layers are more vulnerable. However, to the best of our knowledge, none of the publicly available platforms provide explicit information about affected layers for Android-related vulnerabilities. %neither on any Android security bulletins nor on CVEDetails web page is explicitly indicated affected Android stack layers. 
Thus, a further analysis is essential to determine them. Note that patched vulnerabilities have references in our dataset. %The idea of identifying the affected Android stack layers was arised by analyzing the vulnerability reference links and their details. 
%Given these references links for each vulnerability, it
They contain links that navigate to the code repositories such as Qualcomm Code Aurora, AOSP Git repository and Linux Patchwork. In this classification, we only investigate AOSP Git repositories. For instance, \texttt{CVE-2016-5131}, published in June 2017, has one reference link that navigates to the AOSP Git repository~\cite{Android_Security_Bulletin_2017_06}. In AOSP Git repository, the followings are published: the branch, the commit id, the changed files, the author, the commit message and commit logs.~\cite{CVE_2016_5131_Git_Repository}. 

The references can be more than one as well. For instance, \texttt{CVE-2018-9440} is published in September 2018 and has two different references~\cite{Android_Security_Bulletin_2018_09}. Although they have the same branch which is \textit{frameworks/av}, they have different commit IDs. Besides, one of the references has one additional changed file than the other one which is \textit{media/libstagefright/httplive/M3UParser.h}~\cite{CVE_2018_9440_First_Git_Repository, CVE_2018_9440_Second_Git_Repository}. Moreover, the branches can also be different among the references that belong to the same patched vulnerability. For instance, \texttt{CVE-2017-0831} is published in November 2017 and has two different references \cite{Android_Security_Bulletin_2017_11}. The first reference has the branch \textit{frameworks/base} while the other one has the branch \textit{packages/apps/Settings}. Given this, the changed files can differ from each other as well.

Another issue we notice is that there might be patched vulnerabilities that do not have any references. According to the information that is published on each Android security bulletin, these patched vulnerabilities are marked with the sign of asterisk \textit{*} since their references are not publicly available.

For our classification method, only a branch name itself is not enough to find which layer is affected by a vulnerability. The reason is that there might be directories on the branch that might point to the different stack layers. For this reason, we also need to check the changed files' directories to achieve more accurate classification. Since the changed files have their own directories, we combine the branch path with the changed file patch to get \textit{full path}.

%Moreover, some of the patched vulnerabilities have more than one references and therefore, have different branches and changed files. In other words, there could be more than one layer being affected even in one particular branch. Since, different directories may relate to different stack layers. Since, different directories may relate to different stack layers.

Aside from the full paths, we also use \textit{component name} and \textbf{category name} to identify the stack layers of patched vulnerabilities. Component name is presented as a column of each patched vulnerabilities. Note that there might be such cases where it does not have the component name. Category name, on the other hand, is the title of each category on the Android security bulletins. For instance, in September 2017~\cite{Android_Security_Bulletin_2017_09}, the first category name is \textit{Framework} which consists of only one patched vulnerability. The second category name is \textit{Libraries} that contains three different patched vulnerabilities. Although the component and category names do not provide too much detailed information, we benefit from them to identify external components and kernel stack layer vulnerabilities.

The following subsections describe the classification methods for each stack layers.

%Another issue we notice is that some patched vulnerabilities do not have any references. According to the information that is published on each Android security bulletin, these vulnerability patches have been marked with the sign of asterisk \textit{*} as their references are not publicly available. As a result, we use the commit ID, the full branch name and all the changed files with their directories to classify the vulnerability patches into stack layers.

\subsection{Kernel Layer}

%The first stack layer that was classified is the kernel layer since it is the most straightforward layer for the classification. Since 
All the kernel layer changes are implemented under the branch name \textit{kernel/}~\cite{Android_Kernel_Stack_Layer, Kernel_git}. Thus, a patched vulnerability can be classified as a kernel layer by looking at this branch name. In case of where a patched vulnerability does not have any references, then we look at its category name and the component name. If either of these names has the name \textit{kernel}, we consider it as a Kernel Layer patched vulnerability. For example, \texttt{CVE-2014-9322}, published in April 2016, has the branch name \textit{kernel/}~\cite{Android_Security_Bulletin_2016_04} which makes it a kernel layer patched vulnerability. On the other hand, \texttt{CVE-2017-0648}, published in June 2017, does not have any references~\cite{Android_Security_Bulletin_2017_06}. However, it has the category name \textit{kernel}. Thus, this patched vulnerability is classified as a kernel layer. As a result, we check the branch name, category name, and component name of the patched vulnerabilities to classify them as kernel layer.

\subsection{Hardware Abstraction Layer}
%Identification of HAL Vulnerabilities}

``Hardware Abstraction Layer (HAL) defines a standard interface for hardware vendors to implement, which enables Android to be agnostic about lower-level driver implementations'' \cite{HAL}. An interface description language (IDL) that determines the interface between users and HAL is called HAL interface definition language (HIDL). This also specifies types and method calls, collected into interfaces and packages~\cite{HIDL}. 
%``HAL interface definition language (HIDL) is an interface description language (IDL) to specify the interface between a HAL and its users. It allows specifying types and method calls, collected into interfaces and packages. More broadly, HIDL is a system for communicating between codebases that may be compiled independently''~\cite{HIDL}. 

Based on the Android source~\cite{HIDL_Interfaces}, if a patched vulnerability has the following branch names, then it can be classified as HAL: \textit{hardware/}, \textit{hidl/} and \textit{hwservicemanager/}. The branches \textit{vendor/} and \textit{device/} can also be counted as HAL-related branches, according to~\cite{HIDL_Interfaces} and~\cite{yaghmour2013embedded} Page 88. Moreover, the branch \textit{system/bt/} also indicates a Bluetooth stack located in HAL~\cite{Android_Bluetooth}.

For instance, \texttt{CVE-2017-0767}, published in September 2017, has two references~\cite{Android_Security_Bulletin_2017_09}. One of them has the branch name \textit{hardware/}. Therefore, we classify this patched vulnerability as HAL.
%The first reference has a different branch name, the second reference is in the branch \textit{hardware/}. 
The second example, \texttt{CVE-2017-0812}, published in October 2017, has the branch name \textit{device/}~\cite{Android_Security_Bulletin_2017_10}. Hence, we also classify this as HAL.

\subsection{External Layer}
%Identification of External Vulnerabilities}

Android uses external libraries that are developed by third-parties. These externals might affect different Android stack layers. However, they need further analyses in order to classify them as one of the Android Stack Layers. Thus, we consider them separately. External-related patched vulnerabilities have their own unique branch name \textit{external/}. Thus, we use this branch name to classify the patched vulnerabilities as an external-related one. It consists of all external libraries such as \textit{chromium-libpac} and \textit{v8}. For example, \texttt{CVE-2018-9490}, published in October 2018, has two references~\cite{Android_Security_Bulletin_2018_10}. Both of them have the branch name \textit{external/}. Therefore, we classify this patched vulnerability as an external-related patched vulnerability.

%External libraries are developed by but are imported into AOSP, like third-party applications.In our analysis, we consider them separately as in order to classify them into Android stack layers, they need further analysis. Since external layer patched vulnerabilities have their own unique directory named \textit{external/}, we use this directory name to classify the external-related patched vulnerabilities. This directory consists of all external libraries such as \textit{chromium-libpac} and \textit{v8}. For example, \texttt{CVE-2018-9490}, published in October 2018 has two references~\cite{Android_Security_Bulletin_2018_10}. Both of them the branch name \textit{external/}. Therefore, we classify it as external-related patched vulnerability. 

\subsection{Applications Layer}
%Identification of Applications Layer Vulnerabilities}

Applications Layer is the top layer on all Android stack layers. It consists of different stock applications such as Calculator and Email that the end user interacts with. Based on~\cite{yaghmour2013embedded} (page 88, Figure 3.2), we consider all the changes implemented in the branch \textit{packages/apps} as application layer patched vulnerabilities. For instance, \texttt{CVE-2018-9501}, published in October 2018,
%as an example of application layer vulnerability, was published in October 2018~\cite{Android_Security_Bulletin_2018_10}. It 
has one reference under the branch name \textit{packages/apps}~\cite{Android_Security_Bulletin_2018_10}.

\subsection{Application Framework Layer}
%Identification of Application Framework Layer Vulnerabilities}

Application framework layer consists of abstractions and APIs in order to provide a communication between the applications and the native libraries \cite{Android_Architecture}. According to~\cite{yaghmour2013embedded} (page 65, Figures 2-4), there are Java-built services under the system servers. Java providers and managers are also under the application frameworks layer~\cite{yaghmour2013embedded} (page 73). Moreover, compatibility test suites (CTS)\footnote{CTS tests the overall functionality including the application layer framework} are also considered as one of the part of this layer. We use the following branches and directories to classify the patched vulnerabilities as application framework layer:\textit{platform/packages/providers}, \textit{platform/frameworks/base/}, \textit{platform/packages/services/}, \textit{platform/frameworks/opt/}, \textit{platform/cts/}, \textit{platform/libcore/}.

The branch \textit{platform/frameworks/base/} shares files which also belong to the native libraries layer. To address this issue, the files that have \textit{.java} file extension are counted as the application framework layer while the others that have \textit{.c} or \textit{.cpp} file extensions are considered as native libraries layer.

For example, \texttt{CVE-2018-9438}, published in August 2018, has \textit{platform/packages/providers} branch ~\cite{Android_Security_Bulletin_2018_08}.
%as an example of having \textit{platform/packages/providers} directory, was published in June 2018 \cite{Android_Security_Bulletin_2018_06}. 
Another example is \texttt{CVE-2018-9493} which is published in October 2018~\cite{Android_Security_Bulletin_2018_10}. It has branch name ~\textit{platform/frameworks/base/}. Therefore, we count these two patched vulnerabilities as application framework layer due to their branch names. 
%CVE-2018-9467, was published in September 2018 \cite{Android_Security_Bulletin_2018_09}, was changed under \textit{platform/libcore/} directory. CVE-2015-3865, on the other hand, was published in August 2015 \cite{Android_Security_Bulletin_2015_08} and changed under \textit{platform/cts/}. All of these examples were counted as application framework layer vulnerability patches in the security bulletins dataset.

\subsection{Native Libraries Layer}
%Identification of Native Libraries Layer Vulnerabilities}

Native libraries consist set of native headers and shared library files in order to provide a solid connection between the upper layers and lower layers ~\cite{androidnamespace_libraries, Android_NDK_Native_APIs}. Audio and Media are two examples that part of native libraries~\cite{Android_Audio,Android_Media} %in this layer. 
The branch \textit{/platform/frameworks/av/} indicates the media related libraries~\cite{Android_Stagefright_Media_Player}. We also count \textit{platform/frameworks/base/} since it is the old version of \textit{platform/frameworks/av/}. Further, the branch \textit{platform/frameworks/native/libs/} consists of several native libraries as well~\cite{yaghmour2013embedded} (page 89). 

We also consider some parts of the Android architecture, such as \textit{Native Daemons} and \textit{Init/Toolbox} as parts of the native libraries layer. Since we analyze the patched vulnerabilities by attaching to the Android stack layers and Android stack layer schema does not explicitly indicate these parts, we consider these parts as native libraries as well. For instance, \texttt{CVE-2018-9491}, published in October 2018, has the branch \textit{platform/frameworks/av}~\cite{Android_Security_Bulletin_2018_10} and  \texttt{CVE-2017-0426}, published in February 2017, has the branch \textit{/platform/system/core/}~\cite{Android_Security_Bulletin_2017_02}. Therefore, we consider both of them as native libraries patched vulnerabilities.

\subsection{Android Runtime}
%Identification of Android Runtime Vulnerabilities}

Android Runtime (ART) is an application runtime environment. Replacing Dalvik, the process virtual machine originally used by Android, ART performs the translation of the application's bytecode into native instructions that are later executed by the device's runtime environment \cite{Android_Runtime}. According to \cite{yaghmour2013embedded} Page 88 Figure 3.2, \textit{/dalvik/} branch is under ART. Besides, with the investigation of the directories, we also see files/directory names \textit{androidruntime} or \textit{android\_runtime} in some references. Therefore, we look at these branches and directories/file names to consider a patched vulnerability as ART. For instance, \texttt{CVE-2015-3865}, published in August 2015, has the directory \textit{android\_runtime} in its reference~\cite{Android_Security_Bulletin_2015_08}. Similarly, \texttt{CVE-2016-3758}, published in July 2016, has the branch \textit{platform/dalvik}~\cite{Android_Security_Bulletin_2016_07}.

In Appendix~\ref{app:directories}, we list all branches and directories that we use for the classification of the vulnerabilities published in Android security bulletins.

\subsection{All Branches and Directories for the Classification of the Patched Vulnerabilities}
\label{app:directories}

\begin{itemize}
    \item \textbf{Kernel Layer Branches and Directories}
    \\ ``platform/kernel/''
    
    \item \textbf{HAL Branches and Directories}
    \\``platform/hardware/'', 
    \\``platform/device/'',
    \\``platform/vendor/''
    \\``platform/system/bt/'',
    \\``platform/system/nfc/'',
    
    \item External-Related Branches and Directories:
    \\ ``platform/external/''

    \item \textbf{Applications Layer Branches and Directories}
    \\ ``platform/packages/apps/''
    
    \item \textbf{Application Framework Layer Branches and Directories}
    \\``platform/packages/providers/'',
    \\``platform/frameworks/base/'',
    \\``platform/packages/services/'',
    \\``platform/frameworks/opt/'',
    \\``platform/cts/'',
    \\``platform/libcore/'',
    \\``platform/frameworks/base/services/'',

    \item \textbf{Native Libraries Layer Branches and Directories}
    \\``platform/frameworks/av/'',
    \\``platform/frameworks/base/lib/'',
    \\``platform/frameworks/native/'',
    \\``platform/frameworks/base/core/'' ,
    \\``platform/frameworks/base/media'' ,
    \\``platform/frameworks/minikin/libs/'',
    \\``platform/system/'',
    \\``platform/frameworks/ex/'',
    \\``platform/bionic/'' ,
    \\``platform/bootable/'' ,

    \item \textbf{Android Runtime Layer Branches and Directories}
    \\``platform/dalvik/'',
    \\``platform/frameworks/base/include/android\_runtime'',
    
\end{itemize}

%%%%%%%%%%%%%%%%%%%%%%%%%%%%%%%%%%%%%%%%%%%%%%%%%%%%%%%%%%%%%%%%%%%%%%%%%%%%%%%%
\end{document}